\documentclass[11pt]{article}
\usepackage[utf8]{inputenc}
\usepackage{amssymb,amsmath,mathrsfs,enumerate}
\usepackage{graphicx,rotate,multicol}
\usepackage{float}
\usepackage{tocloft}
\usepackage{subfig}
\usepackage[margin=10pt,labelfont=bf]{caption}
\usepackage{cite}
\usepackage{soul}
\usepackage[colorlinks=true,
linkcolor=blue,
urlcolor=blue,
citecolor=red]{hyperref}
\usepackage{multirow}
\usepackage[table]{xcolor}

\makeatletter
\let\Hy@linktoc\Hy@linktoc@page
\makeatother

\usepackage{color}
\definecolor{ourcolor}{rgb}{0.7, 0.25, 0.05}

\let\tilde=\widetilde
\let\hat=\widehat
\let\bar=\overline

\def \order(#1){{\mathcal O} \left(#1 \right)}

\def\l {\lambda}

\def\bar {\overline}
\def\be {\begin{equation}}
	\def\ee {\end{equation}}
\def\beq {\begin{equation}}
	\def\eeq {\end{equation}}
\def\bea {\begin{eqnarray}}
	\def\eea {\end{eqnarray}}
\newcommand{\besub}{\begin{subequations}}
\newcommand{\eesub}{\end{subequations}}

\evensidemargin=\oddsidemargin
\allowdisplaybreaks

\title{\bf Radiative symmetry breaking in a gauged Zee-Babu model and its gravitational wave imprints}

\author {\bf   Indra Kumar Banerjee,$^{a,}$\footnote{indrab@iiserbpr.ac.in} 
	\hspace{4pt}   Nabarun Chakrabarty,$^{b,}$\footnote{nabarun.chakrabarty@visva-bharati.ac.in} 
	\hspace{4pt}   Ujjal Kumar Dey$^{a,}$\footnote{ujjal@iiserbpr.ac.in} \\[10pt]
	\small\em $^a$Department of Physical Sciences, Indian Institute of Science
	Education and Research Berhampur, \\ \small\em Ganjam, Odisha 760003, India \\
	\small\em $^b$Department of Physics, Siksha Bhavana, Visva-Bharati,\\
	\small\em Santiniketan, West Bengal 731235, India
}

\date{}

\begin{document}
	
\maketitle	

\begin{abstract} 
We construct a classically scale invariant version of the Zee-Babu model governed by an $U(1)_{B-L}$ gauge symmetry wherein three right handed neutrinos with identical gauge charges are present. A $\mathbb{Z}_2$ symmetry is additionally imposed such that the lightest right handed neutrino becomes a dark matter candidate. A spontaneous breakdown of the $U(1)_{B-L}$ gauge group is triggered radiatively through renormalisation group effects and the dimensionful parameters thus emerging are proportional to the corresponding breaking scale $v_{BL}$. We demonstrate in this study how the same $v_{BL}$ controls the dynamics of neutrino mass generation, lepton flavour violation and dark matter phenomenology. It is revealed that the scenario can simultaneously accommodate the observed neutrino masses and mixings, an appropriately low lepton flavour violation and the observed dark matter relic density for 10 TeV $\lesssim v_{BL} \lesssim$ 55 TeV. In addition, the very radiative nature of the set-up signals a strong first order phase transition in the presence of a non-zero temperature. Stochastic gravitational waves stemming from this phase transition are within the reach of detectors such as LISA and BBO. The scenario therefore emerges as a concrete platform to test classical scale invariance that is tied to neutrino masses and dark matter, through gravitational waves. 
\end{abstract} 


\newpage

\hrule \hrule
\tableofcontents
\vskip 10pt
\hrule \hrule


\section{Introduction}

Despite the success of the Standard Model (SM), several unresolved issues on both theoretical and experimental fronts keep advocating for additional dynamics. One such area is the SM's inability to predict non-zero masses for neutrinos and the observed pattern of neutrino-mixing. Massive neutrinos provide concrete hints of physics beyond the SM (BSM) and many theoretical possibilities have been proposed to account for the lightness of neutrinos (see \cite{King:2003jb, Altarelli:2004za, Mohapatra:2006gs, GonzalezGarcia:2007ib} for some reviews). An elegant framework to generate non-zero masses is the Zee-Babu model \cite{Zee:1985rj, Zee:1985id, Babu:1988ki} that features a singly charged scalar and a doubly charged scalar over and above the SM fields. Neutrino masses are generated at the two-loop level and are proportional to the Yukawa couplings of the new scalars and inversely proportional to the square of their masses. For fixed Yukawas, the two-loop dynamics dictates that the additional scalars are lighter than what they would be in a typical tree-level neutrino mass framework, thereby bringing down their masses within the purview of present and future colliders.   
On another front, around $27\%$ of the total mass energy budget of the universe is found to consist of an hitherto unobserved non-luminous, non-baryonic matter called dark matter (DM) \cite{2009GReGr..41..207Z, Rubin:1970zza,Clowe_2006}. Denoting the corresponding density parameter as $\Omega_{\text{DM}}$, one has \cite{Planck:2018vyg}
\bea
\Omega_{\text{DM}}h^2 = 0.120 \pm 0.001,
\eea
where $h$ = Hubble parameter/(100 km/s/Mpc). If the dark matter is hypothesized to be an elementary particle, then there is no candidate for the same in the SM. While a host of particle DM models with different DM production mechanisms exist in the literature, the Weakly Interacting Massive Particle (WIMP) is the most widely studied (some reviews are \cite{Queiroz:2017kxt,Roszkowski:2017nbc,Arcadi:2024ukq}). Here, the interaction strengths involving the DM and bath particles are in the \emph{weak} ballpark as a result of which DM \emph{freezes-out} from thermal equilibrium thereby generating the relic density. However, the very fact that DM interaction strengths are sizeable has cornered most minimal
WIMP scenarios in view of the null results from direct detection experiments such as XENON-1T \cite{XENON1T:2017pvk}, PANDA-X \cite{PandaX4T:2022fxy} and LUX-ZEPPELIN \cite{LZ:2022ufs} more so for a DM mass in the [10 GeV, 1 TeV] range. The only shot at salvaging such models is when the dark matter mass is outside the aforesaid range.
In recent years, first-order phase transitions (FOPTs) in scalar extensions of the Standard Model have emerged as a promising source of stochastic gravitational waves (GWs) potentially observable by forthcoming space-based interferometers. While the presence of additional bosonic degrees of freedom can accommodate a strong FOPT, the transition dynamics can be significantly enriched in scenarios where symmetry breaking occurs radiatively, leading to the dynamical generation of a new scale. In this context, the possibility that such a dynamically generated scale is directly correlated with the neutrino mass scale is particularly intriguing, and the cosmological implications of this connection remain comparatively unexplored.
Motivated by these considerations, we construct a classically scale-invariant extension of the well known Zee-Babu model \cite{Zee:1985rj, Zee:1985id, Babu:1988ki} by endowing it with an $U(1)_{B-L}$ gauge symmetry. Anomaly cancellation in this framework requires the inclusion of three right-handed neutrinos (RHNs), which we take to be odd under a discrete  $\mathbb{Z}_2$ symmetry thereby rendering the lightest RHN a potential dark matter (DM) candidate.
The $U(1)_{B-L}$ is broken radiatively when the 
renormalisation group (RG)-improved scalar potential develops a minimum at a scale $v_{BL}$. An immediate observation is that the crucial scalar trilinear interaction responsible for neutrino mass in the minimal Zee-Babu case is now generated from a quartic interaction after symmetry breaking. Since the same $v_{BL}$ controls the size of neutrino masses while being subject to lower bounds from collider searches, the model naturally leads to a nontrivial interplay between neutrino phenomenology and symmetry-breaking dynamics. The same connection is also expected in the case of the $\ell_i \to \ell_j \ell_k \ell_l $ and $\ell_i \to \ell_j \gamma$ lepton flavor violating (LFV) processes mediated by the Zee-Babu scalars. In addition, the presence of the $U(1)_{B-L}$ sector opens up new annihilation channels for the dark matter candidate, thereby modifying the relic abundance relative to the minimal Zee–Babu scenario. In addition, incorporating $T \neq 0$ corrections in the RG-improved potential can give rise to strong FOPTs thereby providing a source of GWs. In all, we aim to see if the observed pattern of neutrino masses, DM relic density, an appropriately low LFV and detectable GWs can be simultaneously accommodated by a judicious choice of the symmetry breaking scale $v_{BL}$ in the otherwise scale invariant scenario. Some aspects of the $U(1)_{B-L}$ extension in the context of dark matter, neutrino, gravitational waves and allied phenomena are discussed in~\cite{Dey:2025pcs, Banerjee:2024fam, Fu:2023nrn, Hasegawa:2019amx}. 
This article is organised as follows. We introduce the scale-invariant Zee-Babu framework in section \ref{model} and discuss the radiative symmetry breaking. Section \ref{low energy} details a fitting of neutrino data in this framework. The same section also translates bounds coming from LFV and colliders to constraints on the model parameters. The DM phenomenology is outlined in section \ref{dm}. FOPT and production of GWs is discussed in section \ref{fopt} and we conclude in section \ref{conclu}. Important formulae are relegated to the Appendix.

\section{Scale-invariant Zee-Babu framework and radiative symmetry breaking}
\label{model}
\begin{figure}[!htb]
\centering
\includegraphics[scale=0.45]{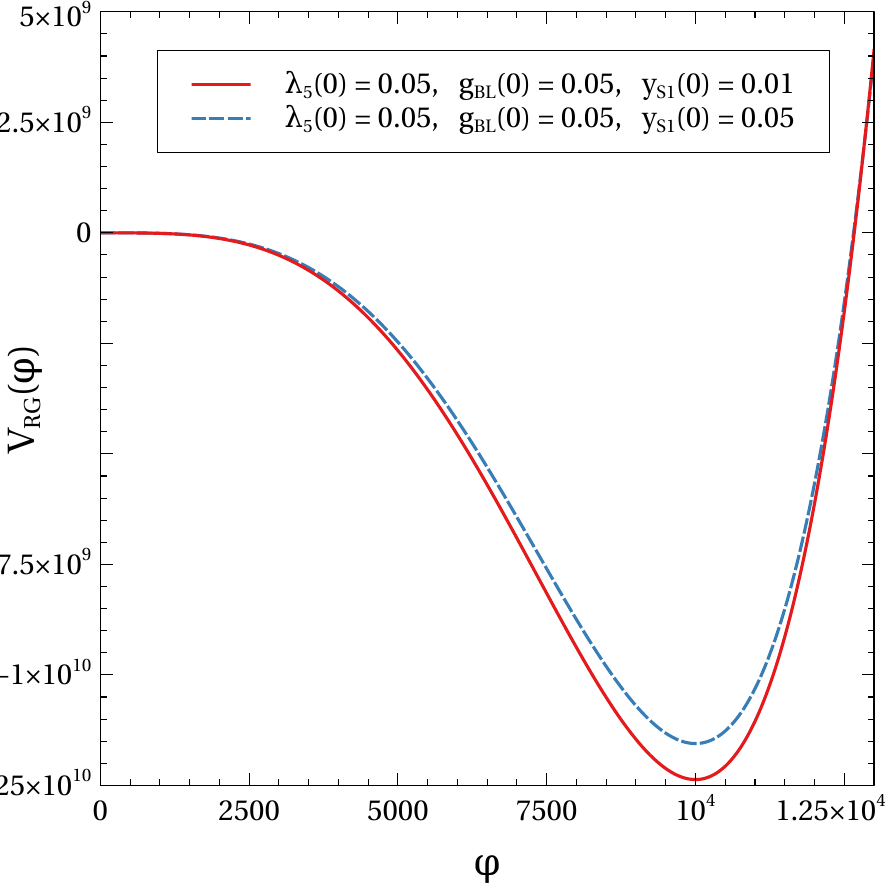} \qquad \qquad
\includegraphics[scale=0.45]{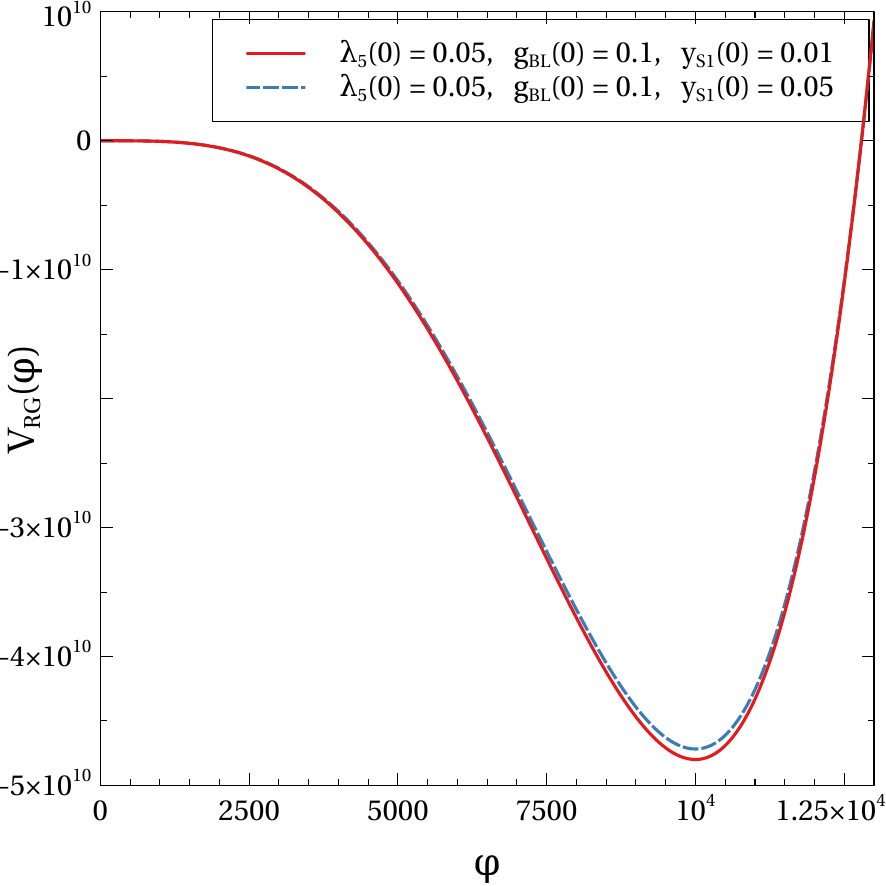}
\caption{Variation of $V_{\text{RG}}(\phi)$ versus $\phi$ for $v_{BL} = 10^4$ GeV and for the choices of $\l_5(0), g_{BL}(0)$ and $y_{1}(0)$ mentioned in the legends. We have further taken $y_{S2}(0)=y_{S3}(0)=2y_{S1}(0)$ in this case.}
\label{fig:VRG}
\end{figure}

We first describe the minimal Zee-Babu model in a nutshell in this section. Over and above the SM fields, this model employs two scalars $h^+$ and $k^{++}$ whose SM quantum numbers are given below. 
\begin{center}
\begin{tabular}{|c| c| c| c|} 
 \hline
 Field & $SU(3)_c$ & $SU(2)_L$ & $U(1)_Y$ \\ [0.5ex] 
 \hline \hline
 $k^{++}$ & 1 & 1 & 2 \\ \hline
 $h^{+}$ & 1 & 1 & 1  \\ \hline
\end{tabular}
\end{center}
The scalar potential is given by,
\begin{align}
\label{eq:VtreeZB1}
V_{\text{tree}} = \mu_H^2 H^\dagger H &+ \mu_h^2 |h^+|^2
 + \mu_k^2 |k^{++}|^2 + \left(\mu_3 k^{++} h^- h^- + \text{h.c.} \right)  + \l_H (H^\dagger H)^2 + \l_h (h^+ h^-)^2 \nonumber \\
 &+ \l_k (k^{++} k^{--})^2 + \l_1 (H^\dagger H) (h^+ h^-) +  \l_2 (H^\dagger H) (k^{++} k^{--}) + \l_6 (h^+ h^-)(k^{++}k^{--}),
\end{align}
where $H = \begin{pmatrix}
G^+ \\
\frac{1}{\sqrt2}(v + h_0 + i G_0)
\end{pmatrix}$ is the SM-like $SU(2)_L$ Higgs doublet and $v$ = 246 GeV is the vacuum expectation value (VEV). Next, the Yukawa interactions involving the BSM fields are,
\bea
-\mathcal{L}^{h,k}_Y =  \Big[ \sum_{i,j =1,2,3} y_1^{ij} \bar{L^c_{Li}}L_{Lj} h^+ + \sum_{i,j =1,2,3} y_2^{ij}\bar{l^c_{Ri}}l_{Rj} k^{++} + \text{h.c.}\Big].
\eea
Assuming the scenario is governed by a global $U(1)_L$, the scalars $h^+$ and $k^{++}$ each carry a charge = $+2$ under the same. It then follows that the trilinear term in the scalar potential violates $U(1)_L$ by 2 unit, opening up a possibility of generation of Majorana masses for the neutrinos. In the context of the Zee-Babu model, this famously occurs
through a two-loop amplitude. The Zee-Babu model is subject to constraints from colliders, non-standard neutrino interactions \cite{Ohlsson:2009vk, Babu:2019mfe} and lepton flavor violation (LFV). Studies on the stability of the electroweak vacuum are performed in \cite{Chao:2012xt, Chun:2014qga} while radiative corrections to couplings are computed in \cite{Kanemura:2014goa}. Collider signatures of the Zee-Babu model have been investigated in \cite{Nebot:2007bc, Schmidt:2014nya, HerreroGarcia:2014hfa, HerreroGarcia:2014proc, Aoki:2011pz, Chun:2012jw, Ruiz:2022ZeeBabu, Jueid:2023ZeeBabuMuon, Crivellin:2018ahb, deMelo:2019doubly}. Electroweak baryogenesis and GWs are looked at in \cite{Phong:2021lea}. Some interesting extensions of this model are \cite{Chen:2022rcv, Okada:2021aoi, Kobayashi:2025hnc, Babu:2024jdw, Chakrabarty:2018qtt}. 
The scalar trilinear operator $k^{++} h^- h^-$ is thus crucial from the perspective of neutrino mass generation in the Zee-Babu set-up. We therefore introduce an additional scalar $S$ and explore the possibility that the aforementioned trilinear operator is generated from a quartic operator via
$S k^{++} h^- h^- \rightarrow \langle S \rangle k^{++} h^- h^-$, where $\langle S \rangle$ denotes the VEV of $S$. This warrants a classical scale invariant approach\footnote{Classical scale-invariance in neutrino mass models was first studied in \cite{Foot:2007ay}.}. Moreover, we stipulate that $S,h^+,k^{++}$ each carry $+2$ units of charge under an $U(1)_{B-L}$ gauge symmetry such that the parent $S k^{++} h^- h^-$ operator stays invariant under the same. The upshot of such a construct is two-fold at this stage. First, the breaking of the $U(1)_L$ is seen as a consequence of a spontaneous breakdown of the $U(1)_{B-L}$. Secondly, the invoked classical scale invariance renders the model economical in terms of parameter count and theoretically attractive. 
A gauged $U(1)_{B-L}$ necessitates introduction of three right handed neutrinos (RHNs) $N_{1,2,3}$ to ensure cancellation of gauge anomalies. A $\mathbb{Z}_2$ symmetry is invoked and $N_i$ are assigned a negative $\mathbb{Z}_2$ charge so that the lightest of them becomes a potential DM candidate. In all, 
the charges of all fields under the $SU(3)_c \times SU(2)_L \times U(1)_Y \times U(1)_{B-L} \times \mathbb{Z}_2$ symmetry\footnote{An $U(1)_{B-L}$ extension of the Zee-Babu model was studied in \cite{Baek:2014sda}. However, the framework was not classically scale invariant and featured a larger field content.} are listed below in table \ref{tab:q_no}.
\begin{table}
\begin{center}
	\rowcolors{2}{gray!20}{white}
\begin{tabular}{|c| c| c| c| c| c|} 
 \hline
 \rowcolor{gray!50}
 Field & $SU(3)_c$ & $SU(2)_L$ & $U(1)_Y$ & $U(1)_{B-L}$ & $\mathbb{Z}_2$ \\ [0.5ex] 
 \hline \hline
 $Q_L$ & 3 & 2 & 1/6 & 1/3 & + \\ \hline
 $L_L$ & 1 & 2 & $-1/2$ & $-1$ & +  \\ \hline
 $u_R, d_R$ & 3 & 1  & 2/3, $-1/3$ & 1/3 & +  \\ \hline
 $l_R$ & 1 & 1 & $-1$ & $-1$ & +  \\ \hline
 $N_{1,2,3}$ & 1 & 1 & 1 & $-1$ & $-$  \\ \hline
 $H$ & 1 & 2 & 1/2 & 0 & +  \\ \hline
 $k^{++}$ & 1 & 1 & 2 & 2 & +  \\ \hline
 $h^{+}$ & 1 & 1 & 1 & 2 & +  \\ \hline
 $S$ &1 &1 &0 & 2 & +  \\ \hline
\end{tabular}
\end{center}
\caption{Quantum numbers of the various fields under the gauge and $\mathbb{Z}_2$ symmetry}
\label{tab:q_no}
\end{table}
In the presence of the additional field $S$, the scalar potential at the tree level can be extended from Eq.~\eqref{eq:VtreeZB1} and is given by,
\begin{align}
V_{\text{tree}} = \l_H (H^\dagger H)^2 &+ \l_h (h^+ h^-)^2 + \l_k (k^{++} k^{--})^2 + \l_S(S^* S)^2 + \l_1 (H^\dagger H) (h^+ h^-) \nonumber \\
&+  \l_2 (H^\dagger H) (k^{++} k^{--}) 
 + \l_3 (H^\dagger H) (S^* S) + \l_4 (h^+ h^-) (S^* S) \nonumber \\
 &+  \l_5 (k^{++} k^{--}) (S^* S) + \l_6 (h^+ h^-)(k^{++}k^{--})
 + \Big[\l_7 S k^{++} h^- h^- + \text{h.c.} \Big]. \label{si_pot}
\end{align}
The multiplets $H$ and $S$ are expressed as
\bea
H = 
\begin{pmatrix}
G^+ \\
\frac{1}{\sqrt{2}}(h_0 + i G_0)
\end{pmatrix},~~~~
S = \frac{1}{\sqrt{2}}(s_0 + i a_0).
\eea
The following Yukawa term can be added with the introduction of the RHNs.
\bea
-\mathcal{L}_Y =  \Big[\sum_{ij=1,2,3} (Y_S)_{ij} \bar{N^c_i}N_j S + \text{h.c.}\Big].
\eea
The Yukawa coupling matrix $Y_S$ can be chosen to be diag$(y_{S1},y_{S2},y_{S3})$ and real without any loss of generality. 
Next, we explore possible radiative breaking of the $U(1)_{B-L}$ symmetry in the direction of $S$. The RG-improved potential for $S = \frac{\phi}{\sqrt2}$ reads
\bea
V_{\text{RG}}(\phi) = \frac{1}{4}\l_S(t)\phi^4. \label{eq:VRG}
\eea
In Eq.~\eqref{eq:VRG}, $\phi$ refers to the background field and $t = \text{log}\big(\phi/\mu \big)$ with $\mu$ denoting the dimensional regularisation scale. The quantity $\l_S(t)$ refers to the quartic coupling $\l_S$ at the corresponding value of $\phi$ and can be obtained by solving the RG equations for this model. Thus, RG equations at a given loop order (say) introduces radiative effects to $V_{\text{RG}}(\phi)$ at that loop order. The dominant contribution typically comes at the one-loop level for the parameter ranges of our interest and therefore we adhere to it accordingly. A complete list of one-loop RG equations for this framework is given in the Appendix. We denote $\langle S \rangle$ by $\frac{1}{\sqrt2}v_{BL}$ in this work and demand that $V_{RG}(\phi)$ has a minimum at $\phi = v_{BL}$. In other words,  
\bea
\frac{d V_{\text{RG}}}{d\phi}|_{\phi=v_{BL}}=0,
\eea
Choosing $\mu = v_{BL}$ leads to
\bea
4 \l_S(0) + \beta_{\l_S}(0) = 0,
\eea
i.e.,
\bea
64\pi^2\l_S(0) + 20 \l_S^2(0) + 2 \l_3^2(0) + \l_4^2(0) + \l_5^2(0) - 48\l_S(0) g^2_{BL}(0) \nonumber \\
+ 8 \l_S(0)\text{Tr}[Y_S^\dagger Y_S](0)
 + 96 g^2_{BL}(0) - \text{Tr}[Y_S^\dagger Y_S Y_S^\dagger Y_S](0) = 0. \label{ls0}
\eea
One numerically solves for $\l_S(0)$ from Eq.~\eqref{ls0} for given values of the other couplings at $t=0$. The one-loop $\beta$-functions for this model are given in the Appendix. Of all model couplings, the RG-improved potential $V_{\text{RG}}(\phi)$ is more sensitive to the ones entering $\beta_{\l_S}$. To illustrate the behaviour of the RG-improved potential, we fix $\l_1(0) = \l_2(0) = \l_3(0) = \l_6(0) = 0.01,~\l_7(0) = 0.7$ and plot $V_{\text{RG}}(\phi)$ versus $\phi$ in Fig.~\ref{fig:VRG} for the shown values of the other couplings.
The following is then inferred. First, $\l_S(0)$ must be negative for $V_{\text{RG}}(\phi)$ to have a minima at $\phi=v_{BL}$. This accordingly constrains the other couplings at $t=0$ through Eq.~\eqref{ls0}. Given the presence of both bosonic as well as fermionic terms in $\beta_{\l_S}$,  Choosing higher values for the former (latter) at $t=0$ accordingly leads to a higher (lower) value of $|\l_S(0)|$. Moreover, the larger (smaller) is $|\l_S(0)|$, the deeper (shallower) is the minima. These inferences are indeed corroborated by Fig.~\ref{fig:VRG}. 
Further, electroweak symmetry breaking (EWSB) occurs when $h_0$ picks up a VEV following which the $SU(2)_L \times U(1)_Y \times U(1)_{B-L} \rightarrow U(1)_Q$ breakdown is complete and therefore $h_0 = v + h_0^\prime,~s_0 = v_{BL} + s_0^\prime$. An $h_0^\prime$-$s_0^\prime$ mixing arises and is controlled by $\l_3(0)$ leading to the mass eigenstates $h,h_2$ through,
\bea
\begin{pmatrix}
h \\
h_2
\end{pmatrix} =
\begin{pmatrix}
\text{cos}\theta & \text{sin}\theta \\
-\text{sin}\theta & \text{cos}\theta
\end{pmatrix}
\begin{pmatrix}
h_0^\prime \\
s_0^\prime
\end{pmatrix},
\eea
where $\theta$ is a mixing angle. Stipulating $v_{BL} \gg v,~\l_3(0) v^2_{BL} = - 2\l_H(0)v^2,~\l_H(0) = m^2_h/(2 v^2)$ ensures a low mixing and an observed Higgs mass $m_h$ = 125 GeV.
We end this section by quoting below the masses for the BSM fields in terms of relevant couplings and VEVs, 
\besub
\bea
m^2_{h^+} &=& \frac{1}{2}(\l_1 v^2 + \l_4 v_{BL}^2),  \\
m^2_{k^{++}} &=& \frac{1}{2}(\l_2 v^2 + \l_5 v_{BL}^2),  \\
m_{Z_{BL}} &=& 2 g_{BL} v_{BL},  \\
m_{N_i} &=& \frac{1}{\sqrt 2}y_{Si} v_{BL}.
\eea
\eesub

\section{Neutrino, lepton flavour violation, and collider constraints}
\label{low energy}
Neutrino masses are generated in this framework at two-loops in the usual Zee-Babu fashion with the trilinear parameter $\mu_3 = \l_7(0)v_{BL}$. The elements of the corresponding mass matrix are given by~\cite{McDonald:2003zj},
\bea
\big(m_\nu\big)_{ij} = \sum_{p,q=1,2,3} 16 \l_7(0) v_{\text{BL}}~y_1^{ip} m_p y_2^{pq}  \mathcal{I}_{pq} m_q y_1^{qj}. \label{numass_formula}
\eea
The function $\tilde{I}_{pq}(r)$ stemming from the two-loop integral can be approximated as,
\bea
\tilde{I}_{pq}(r) &=& \frac{\pi^2}{3 (16\pi^2)^2} \frac{\delta_{pq}}{M^2}g(r),
\eea
with $M$ = max($m_{h^+},m_{k^{++}}$) and 
$g(r) = 1 + \frac{3}{\pi^2}\left(\text{ln}^2(r) - 1\right)$ for $r > 1$ \cite{McDonald:2003zj}. 
Eq.~\eqref{numass_formula} can be expressed in the matrix form as,
\bea
m_{\nu} &=& Y_1^T f Y_1,
\eea
where $f_{ij} = 16 \l_7(0) v_{BL} m_i m_j I_{ij}y_2^{ij}$. The  Pontecorvo-Maki-Nakagawa-Sakata (PMNS) matrix $U$ brings $m_{\nu}$ to a diagonal form through $\hat m_\nu = U^T m_{\nu} U$ with $\hat m_\nu$ = diag($m_{\nu_1},m_{\nu_2},m_{\nu_3}$). Since $\mu_3,~m_{h^+},~m_{k^{++}} \sim v_{BL}$ in this scale invariant set-up, the neutrino mass scale is approximately given by,  
\bea
m_{\nu} \sim \Big(\frac{1}{16 \pi^2}\Big)^2 \Big(\frac{\l_7 a^2 b}{\l_5}\Big) \Big(\frac{m^2_{\text{lepton}}}{v_{BL}}\Big). \label{mnu_appr}
\eea
In Eq.~\eqref{mnu_appr}, $a$ and $b$ are used to denote the orders of magnitude of $y_1^{ij}$ and $y_2^{ij}$. The light neutrino mass scale is therefore dynamically generated from $v_{BL}$. This is a crucial difference from the minimal Zee-Babu model in which case $\mu_3$ and the Zee-Babu masses are \emph{a priori} uncorrelated. Having said so, we now look to fit the neutrino data more accurately in this scenario. Employing the Casas-Ibarra parameterisation \cite{Casas:2001sr} for Majorana neutrino masses, one writes
\bea
Y_1 = V^* f^{-1/2} R~(\hat{m_\nu})^{1/2}~U^\dagger, \label{ci}
\eea
where $V$ diagonalises $f$ through $V^\dagger f V = \hat f$. One notes that $V = I_{3 \times 3}$ here since $f$ is diagonal. More specifically, for 
$m_{h^{++}} > m_{k^+}$, 
\bea
f \sim \frac{32}{(16\pi^2)^2}\Big(\frac{\l_7}{\l_5}\Big)\Big(\frac{1}{v_{BL}}\Big)~\text{diag}(y_2^{ee}m_e^2,y_2^{\mu\mu}m_\mu^2, y_2^{\tau\tau}m_\tau^2).
\eea
In addition, $R$ is an arbitrary $3 \times 3$ complex orthogonal matrix. The PMNS matrix has the form
\bea
U =
\begin{pmatrix}
c_{12} c_{13} &
s_{12} c_{13} &
s_{13} e^{-i\delta} \\

- s_{12} c_{23} - c_{12} s_{23} s_{13} e^{i\delta} &
\;\; c_{12} c_{23} - s_{12} s_{23} s_{13} e^{i\delta} &
s_{23} c_{13} \\

s_{12} s_{23} - c_{12} c_{23} s_{13} e^{i\delta} &
- c_{12} s_{23} - s_{12} c_{23} s_{13} e^{i\delta} &
c_{23} c_{13},
\end{pmatrix}
\eea
with the shorthand $s_{ij}(c_{ij}) = \text{sin}\theta_{ij}(\text{cos}\theta_{ij})$.
Choosing Normal Hierarchy (NH), we take the neutrino oscillation parameters  as,
$\theta_{12}=33.68^\circ,~\theta_{23}=48.5^\circ,~\theta_{13} = 8.52^\circ,~\delta_{CP}=177^\circ,~\Delta m^2_{21}=7.5 \times 10^{-5}~\text{eV}^2,~\Delta m^2_{31}=2.5 \times 10^{-3}~\text{eV}^2$ from the \texttt{NuFit-6.0} analysis~\cite{Esteban:2024eli}.
%

%
The Zee-Babu scalars $h^+$ and $k^{++}$  mediate $\ell_i \to \ell_j \ell_k \bar{\ell}_l$ and $\ell_i \to \ell_j \gamma$ LFV processes. Upper bounds on the corresponding branching ratios therefore accordingly constrain the size of $y_1^{ij}$ and $y_2^{ij}$. We refer to \cite{Herrero-Garcia:2014hfa} for a list of such bounds and refrain from discussing this here in detail for brevity. For instance, the most stringent limits come from $\mu \to e e \bar{e}$ \cite{SINDRUM:1987nra} and $\mu \to e \gamma$ \cite{MEG:2016leq}, and, are seen to lead to \cite{Herrero-Garcia:2014hfa},
\besub
\begin{gather}
|y_2^{ee}(y_2^{e\mu})^*| < 2.3 \times 10^{-5}\Big(m_{k^{++}}/\text{TeV}\Big)^2, \label{lfv1}\\
r^2 |(y_1^{e\tau})^*y_1^{\mu\tau}|^2 + 16 |(y_2^{ee})^*y_2^{e\mu} + (y_2^{e\mu})^*y_2^{\mu\mu} + (y_2^{e\tau})^*y_2^{\mu\tau}|^2 < 1.6 \times 10^{-6}(m_{k^{++}}/\text{TeV})^4. \label{lfv2}
\end{gather}
\eesub
In this study, the matrix $y_1$ is reconstructed for a fixed $y_2$ and $R$ using Eq.~\eqref{ci}. 
\begin{figure}[t]
	\centering
	\includegraphics[scale=0.45]{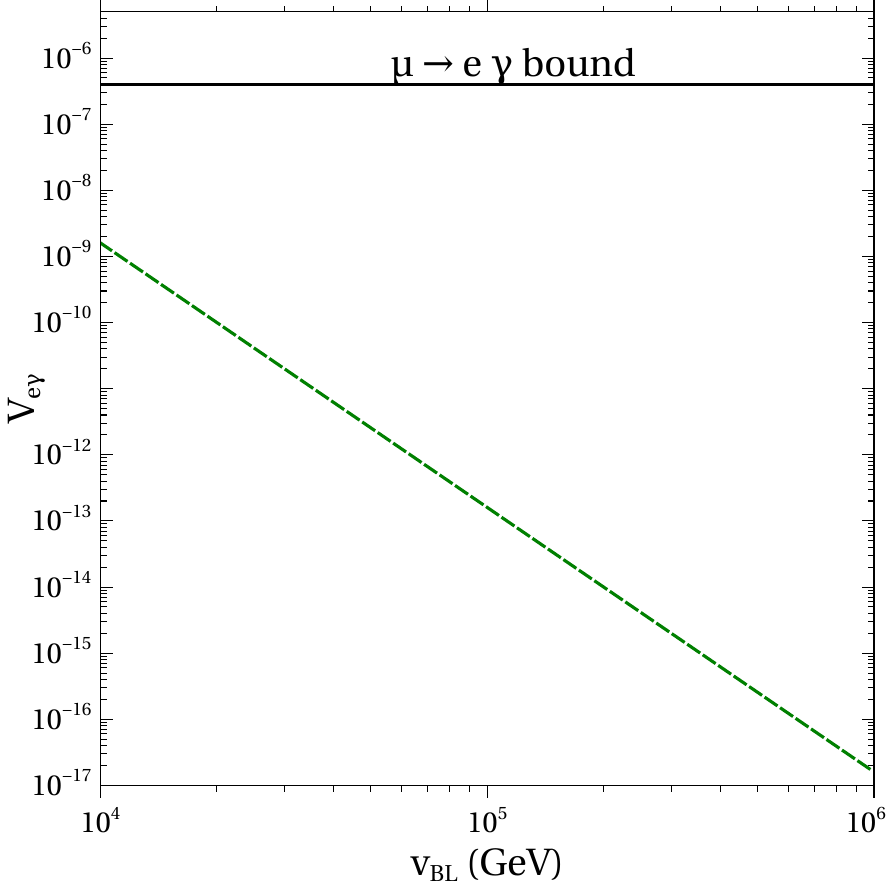}
	\caption{Lepton flavour violation constraints in terms of the dimensionless parameter $V_{e\gamma}$ (see Eq.~\eqref{eq:Vegam}).}
	\label{lfv}
\end{figure}
We next discuss possible constraints from colliders. A non-observation of a $Z_{BL}$ gauge boson at LEP-II enforces $m_{Z_{BL}}/g_{BL} \gtrsim$ 7.1 TeV \cite{Cacciapaglia:2006pk} in this framework which ultimately translates to $v_{BL} \gtrsim$ 3.55 TeV. We adopt a more conservative approach and adhere to $v_{BL} \geq $ 10 TeV in view of tighter bounds in the future. A LEP exclusion limit also reads $m_{h^+} >$ 100 GeV leading to $v_{BL} \gtrsim \frac{142}{\sqrt{\l_4}}$ GeV for $\l_4$ and $\l_1$ of comparable magnitudes. A more stringent exclusion bound of $m_{k^{++}} > $ 600 GeV applies to this setup from the non-observation of the $k^{++} \to \ell^+ \ell^+$ decay mode at the LHC translating to $v_{BL} \gtrsim \frac{142}{\sqrt{\l_5}}$ GeV for $\l_5$ and $\l_2$ of comparable magnitudes.
The bounds stated in Eqs.~(\ref{lfv1})-(\ref{lfv2}) in the present scale-invariant framework can be interpreted on bounds on two dimensionless variables $V_{3e}$ and $V_{e\gamma}$. That is,
\besub
\begin{gather}
V_{e\gamma} \equiv \frac{|y_2^{ee}(y_2^{e\mu})^*|}{\l_5} \frac{1}{(v_{BL}/\text{TeV})^2} < 1.15 \times 10^{-5}, 
\label{eq:Vegam} \\
V_{3e} \equiv \frac{r^2 |(y_1^{e\tau})^*y_1^{\mu\tau}|^2 + 16 |(y_2^{ee})^*y_2^{e\mu} + (y_2^{e\mu})^*y_2^{\mu\mu} + (y_2^{e\tau})^*y_2^{\mu\tau}|^2}{\l_5^2} \frac{1}{(v_{BL}/\text{TeV})^4} < 4 \times 10^{-7}.
\label{eq:V3e}
\end{gather}
\eesub
We observe that choosing $|y_2^{ij}| \gtrsim 10^{-2}$ for $\l_5 \sim 0.1$ and $v_{BL} >$ 10 TeV suffices to bypass all $l_i \to l_j \gamma$ bounds.
Therefore we choose $Y_2 = \begin{pmatrix}
1 & 1 & 1 \\
1 & 1 & 1 \\
1 & 1 & 1.
\end{pmatrix} \times 10^{-2},~\l_4 = 0.1,~\l_5 = 0.3$ and study the variation of $V_{e \gamma}$ in Fig.~\ref{lfv}.

It is therefore inferred from Fig.~\ref{lfv} that the $v_{BL} \gtrsim $ 10 TeV predicts LFV below the current limits while accounting for the observed neutrino masses and mixings.  
\begin{figure}[!htbp]
	\centering
	\includegraphics[scale=0.32]{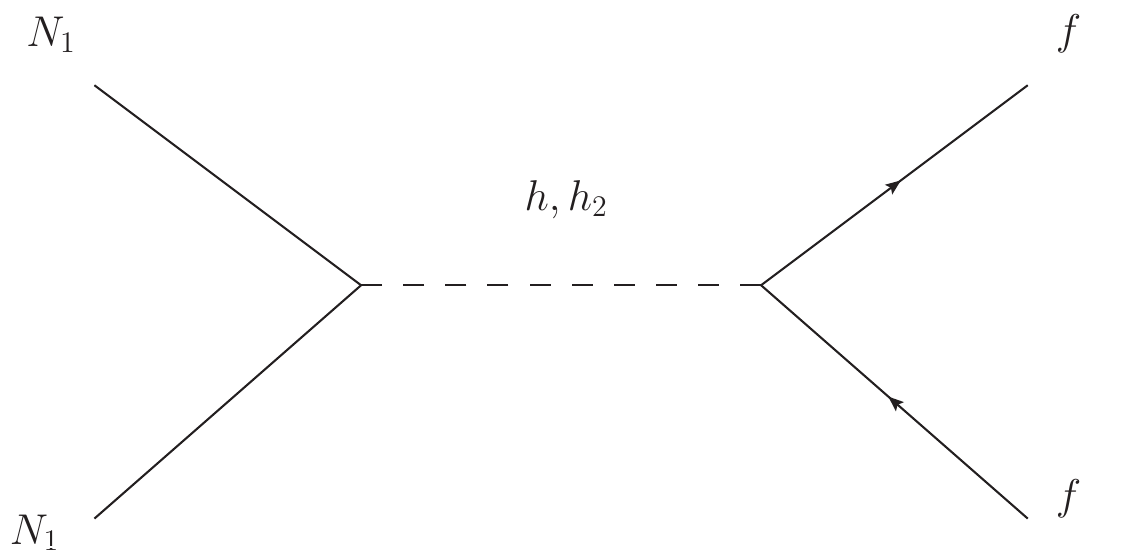}~~~
	\includegraphics[scale=0.32]{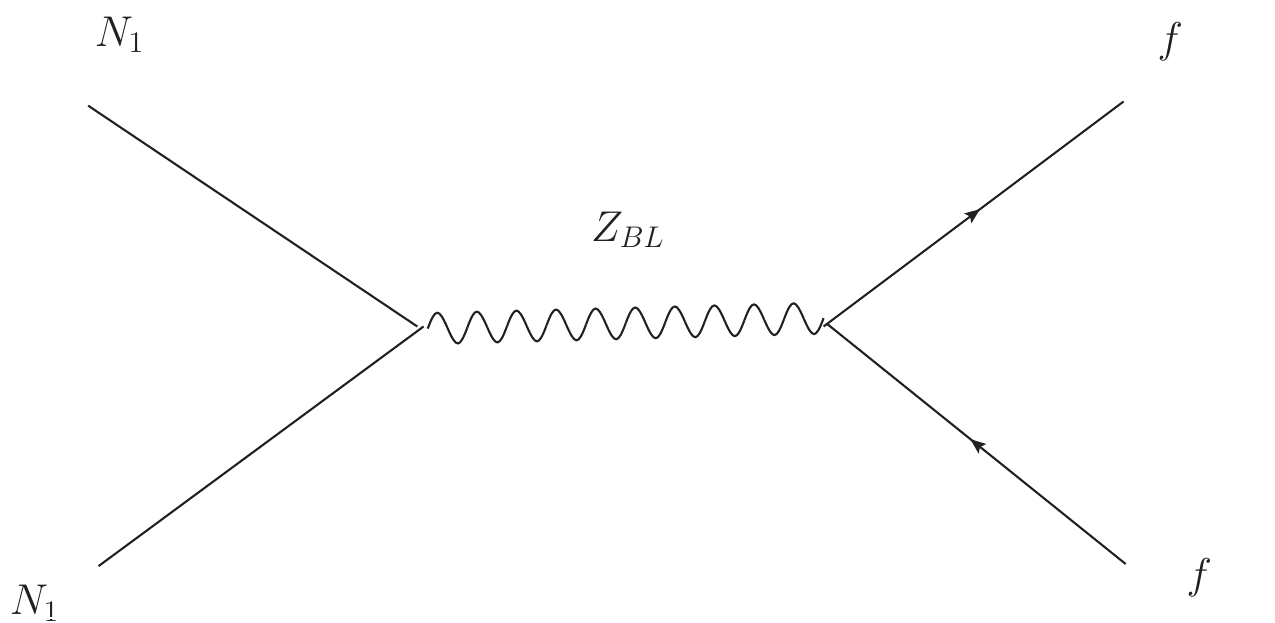}\\
	\includegraphics[scale=0.32]{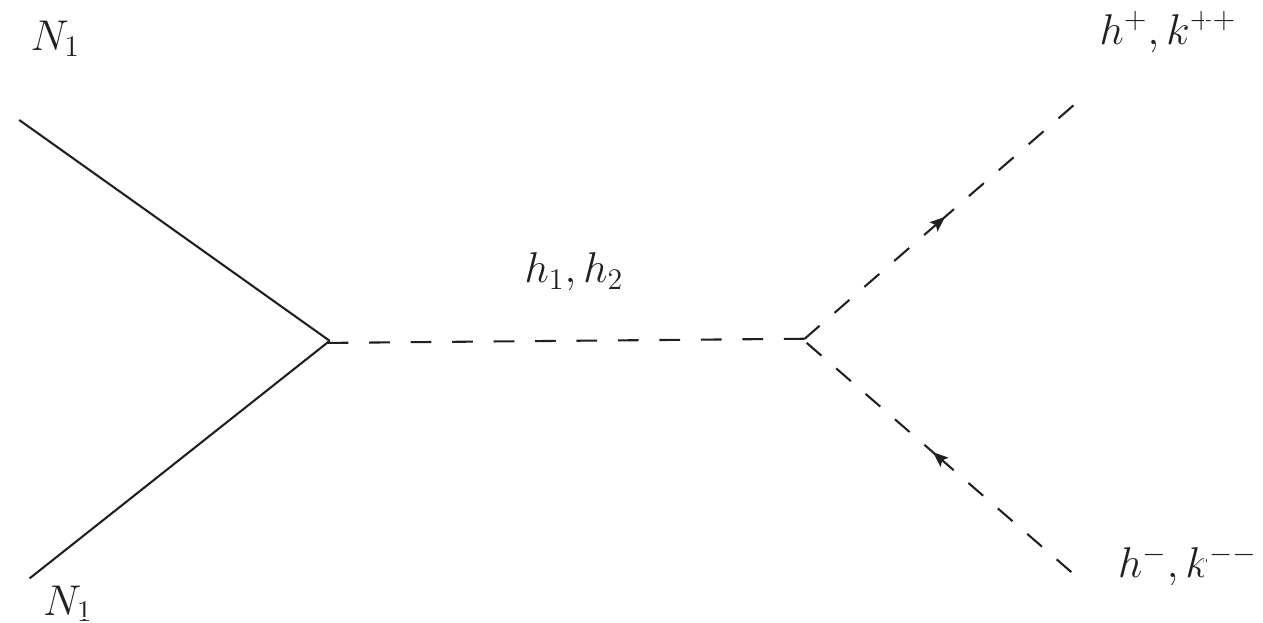}~~~
	\includegraphics[scale=0.32]{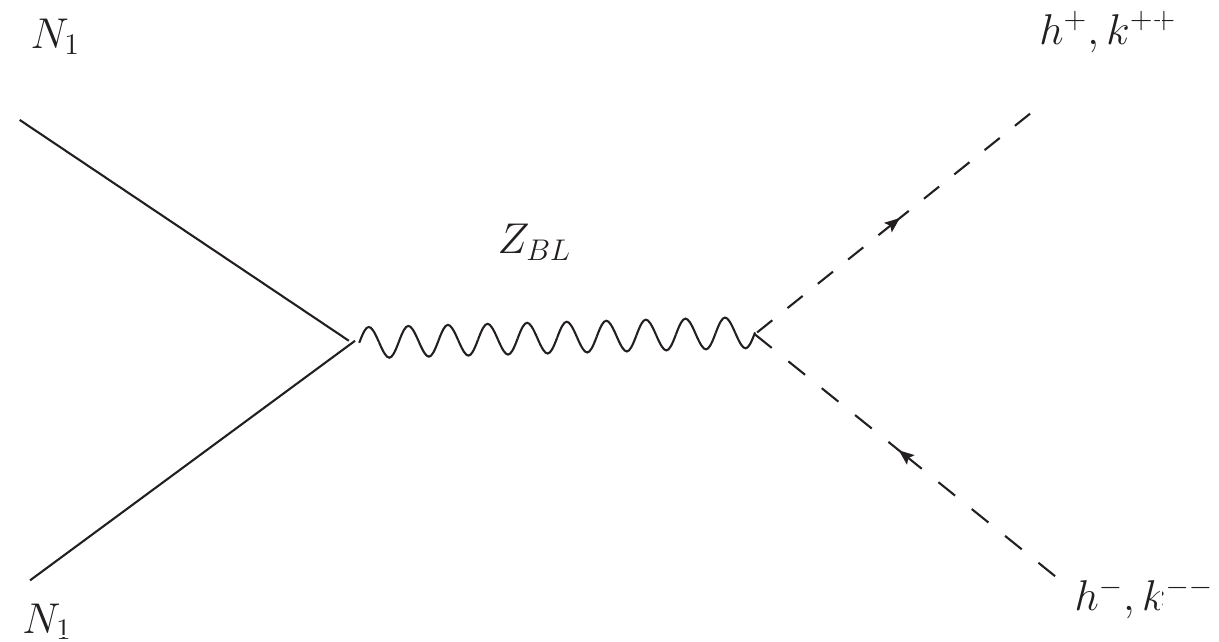}\\
	\includegraphics[scale=0.32]{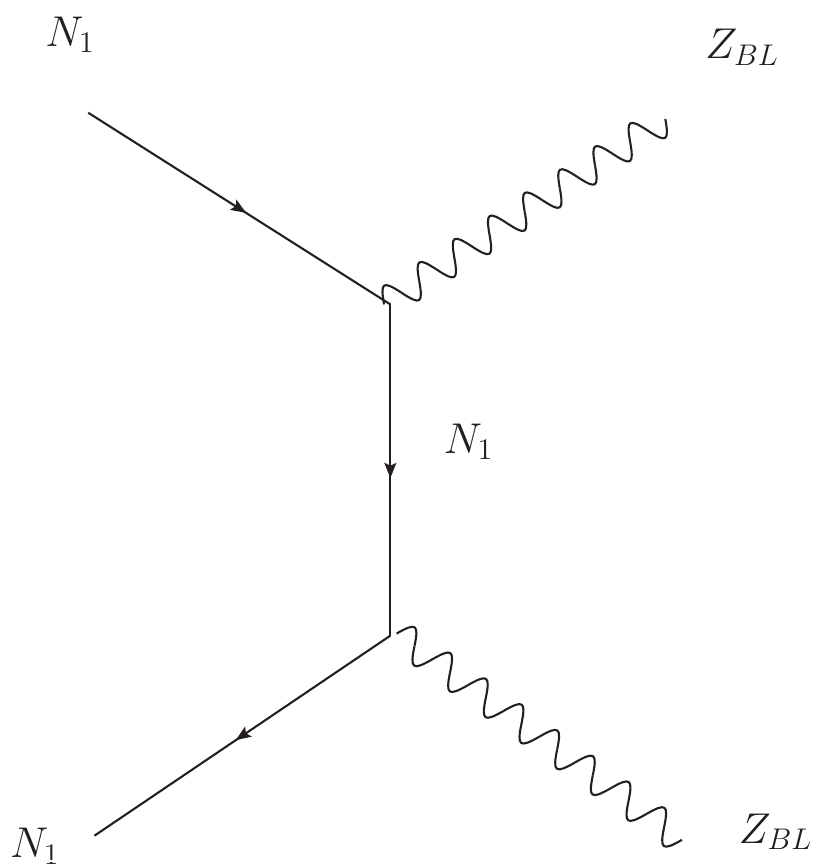}~~~
	\includegraphics[scale=0.32]{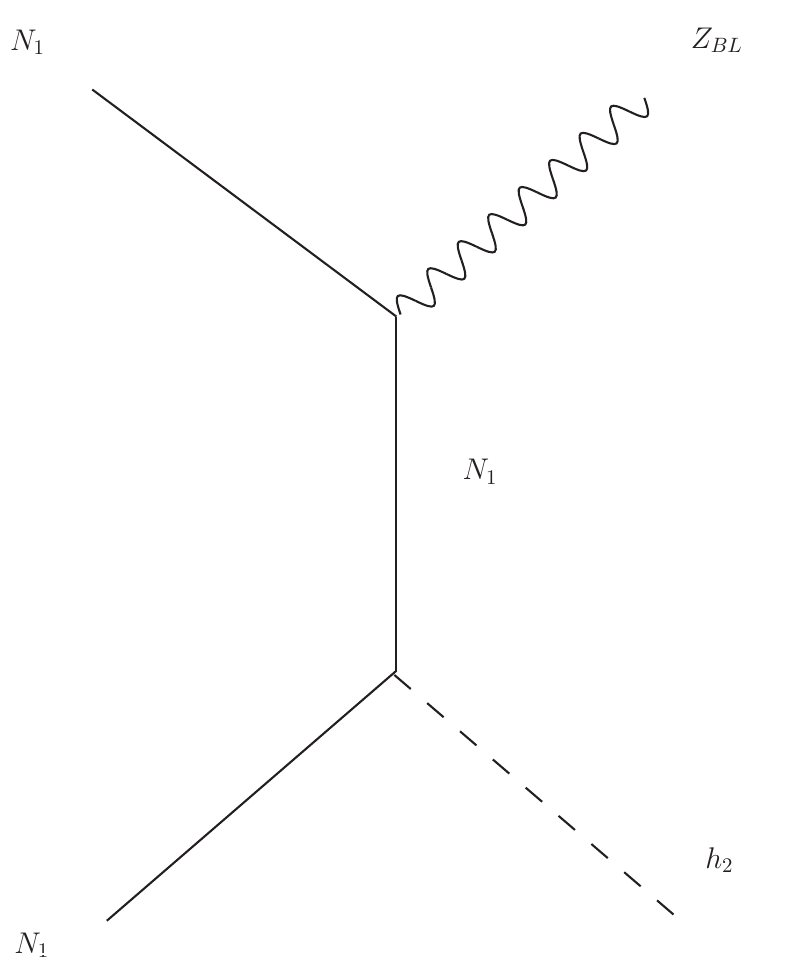}\\
	\includegraphics[scale=0.32]{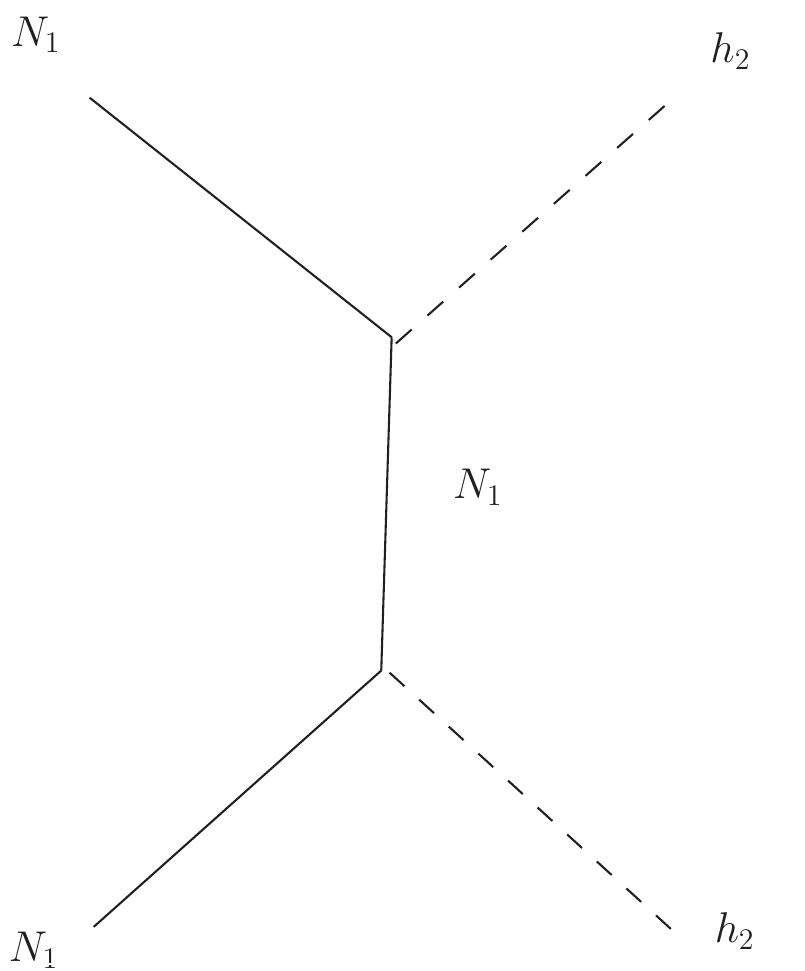}
	\caption{The DM annihilation channels in this model.}
	\label{anni}
\end{figure}
%

\section{Dark matter phenomenology}
\label{dm}
The dark matter phenomenology in the present framework is qualitatively similar to that of minimal $U(1)_{B-L}$ case. We choose $N_1$ as the DM candidate by ensuring that $y_{S1} < y_{S2},y_{S3}$. The DM relic density is dictated by the $N_1 \bar{N_1}$ annihilations to the particles in the thermal bath as shown in Fig.~\ref{anni}.

\begin{figure}[t]
	\centering
	\includegraphics[scale=0.48]{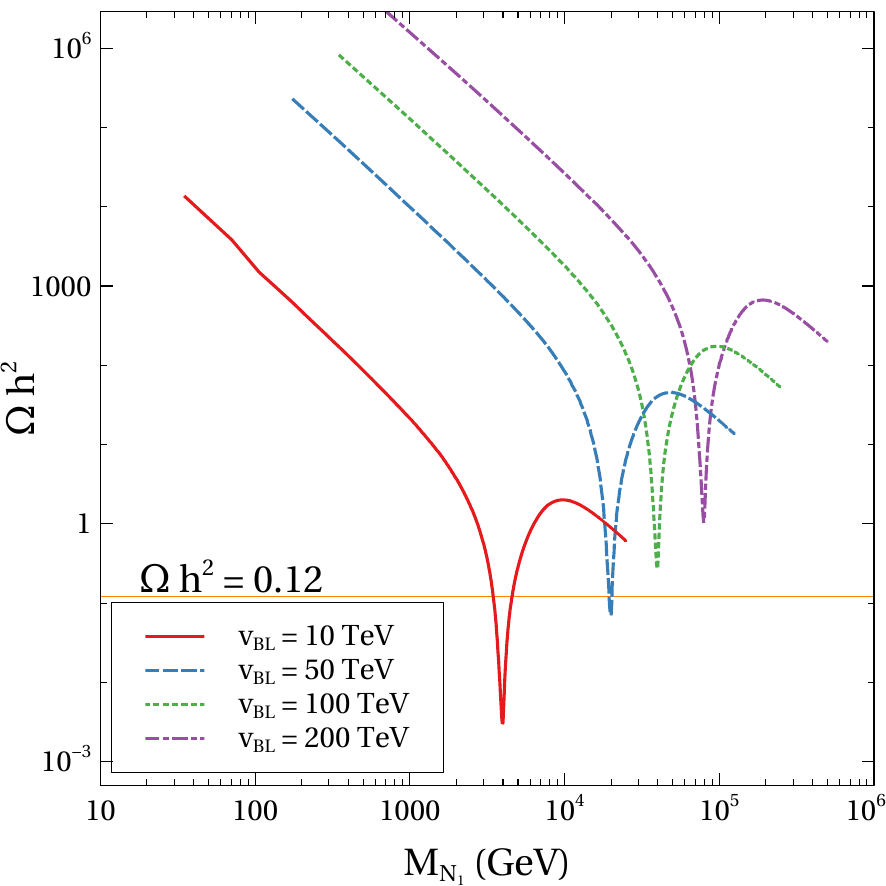}\qquad \qquad
	\includegraphics[scale=0.48]{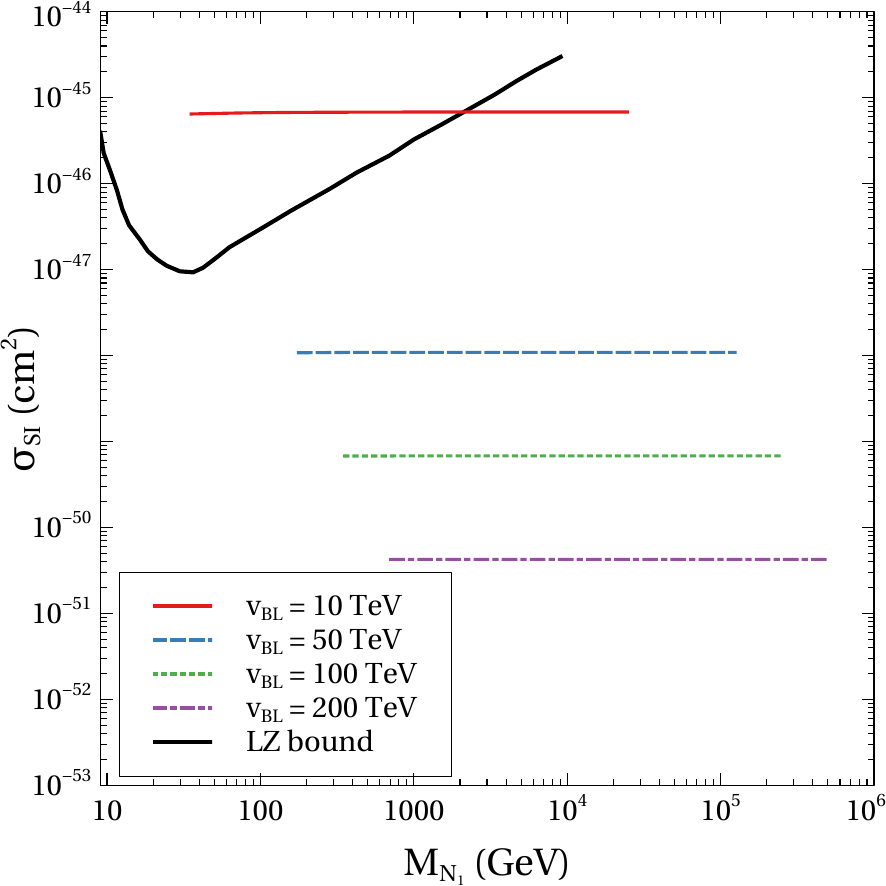}
	\caption{(Left) Variation of $\Omega h^2$ with $M_{N_1}$ for different values of $v_{BL}$. The values of the other model parameters are mentioned in the text. The horizontal line denotes the Planck central value. (Right) Dependence of spin-independent direct detection cross section on the $M_{N_1}$ for different values of $v_{BL}$. The black solid curve represents the bound from LUX-ZEPPELIN experiment.}
	\label{relic}
\end{figure}

It is pointed out that $N_1 \bar{N_1} \to h^+ h^-,~k^{++} k^{--}$ annihilations can possibly occur in this framework that are absent in the minimal $U(1)_{B-L}$ case. We implement the model in \texttt{LanHEP} \cite{Semenov:2008jy}. A compatible output is then fed into the
publicly available tool \texttt{micrOMEGAs} \cite{Belanger:2014vza} to compute the relic density of $N_1$. 
We work in a rather conservative $v_{BL} > $ 10 TeV region in this study in view of the non-observation bound from LHC. The left panel of Fig.~\ref{relic} displays the variation of the relic density with the DM mass for $v_{BL}$ = 10 TeV, 50 TeV, 100 TeV and 200 TeV. As for the values of the other model parameters, it must be noted that the annihilation cross sections have no or negligible dependence on the scalar couplings $\l_H,\l_h,\l_k,\l_1,\l_2,\l_6,\l_7$. Therefore we choose $\l_{h}(0) =\l_{k}(0) = \l_1(0)=\l_2(0)=\l_6(0)=\l_7(0)=0.01$\footnote{The values of the independent parameters are assigned at the breaking scale $v_{BL}$ in our scale-invariant approach.}. We also take $g_{BL}(0) = 0.4$ given that the annihilations are dominated by $Z_{BL}$-mediation and hence the dependence on the gauge coupling is also mild. Finally, we take $3\l_4(0)=\l_5(0) =0.3,~\l_S(0) = -3 \times 10^{-5}$ and vary $0.005 < y_{S1} < \sqrt{4\pi}$ while maintaining $M_{N_1}$ to be the lightest RHN.
Given the dominant $Z_{BL}$ mediation,
resonance dips are observed in $\Omega h^2$ for $2 M_{N_1} \simeq M_{Z_{BL}}$, that is, for $y_{S1}(0) \simeq \sqrt{2} g_{BL}(0)$. We also report an up to $\simeq 20\%$ contribution to the total annihilation cross section from the $h^+h^-,~k^{++}k^{--}$ final states. 
DM-nucleon scattering processes are triggered via $t$-channel mediation of $Z_{BL}$ and are $\propto 1/v^4_{BL}$. Currently the most stringent bound on the spin-independent direct detection
cross section comes from the LUX-ZEPPELIN (LZ) experiment. The cross sections and the LZ bound are both shown in the right panel of Fig.~\ref{relic}. An inspection of both its panels suggests that
stipulating that $N_1$ accounts entirely for the observed DM, and at the same time, predicts direct detection cross section below the latest bound implies that the symmetry breaking scale is bounded from both ends, i.e., $10~\text{TeV} < v_{BL} < 55~\text{TeV}$. Such a constraint on $v_{BL}$ is thus stronger than what is obtained from LFV alone. This point is a key takeaway of our analysis as a result of which we obtain a window of $v_{BL}$ to probe via GWs.

\section{First order phase transition and gravitational waves}\label{fopt}
We first discuss of possibility of FOPT in the present set-up. The thermal potential reads,
\bea
V_T(\phi,T) &=& \sum_{b=h,G^0,G^+,h^+,k^{++},S} n_b J_B\bigg(\frac{M^2_b(\phi)}{T^2} \bigg) + \sum_{f=N_1,N_2,N_3} n_f J_F\bigg(\frac{M^2_f(\phi)}{T^2} \bigg),
\eea
where, the various field-dependent masses in this framework are given by,
\besub
\bea
M^2_{h_0}(\phi) = M^2_{G_0}(\phi) = M^2_{G^+}(\phi) &=& \frac{\l_3}{2}\phi^2, \\
M^2_{h^+}(\phi) = \frac{\l_4}{2}\phi^2,~~M^2_{k^{++}}(\phi) &=& \frac{\l_5}{2}\phi^2, \\
M^2_{s_0}(\phi) = \frac{3\l_S}{2}\phi^2,~~M^2_{a_0}(\phi) &=& \frac{\l_S}{2}\phi^2, \\
M^2_{Z_{BL}}(\phi) = 4 g^2_{BL}\phi^2,~~M^2_{N_i}(\phi) &=& \frac{y^2_{Si}}{2}\phi^2.
\eea 
\eesub
the degrees of freedom of the various bosons and fermions in the model are $n_h = n_{G^0}=n_S=1,~n_{G^+}=n_{h^+}=n_{k^{++}}=2,~n_{N_i} = -4$. The functions $J_{B,F}(x)$ are defined as
\bea
J_{B,F}(x) &=& \int_0^\infty dx~y^2 ~\text{log}\left[1 \mp e^{-\sqrt{y^2 + x}}\right].
\eea 
We also perform resummation in order to incorporate  infra-red (IR) effects. The Arnold-Espinosa technique is adopted in this work in which the following daisy term is added to the thermal potential~\cite{Arnold:1992rz},
\bea
V_{\text{Daisy}}({\phi,T}) &=& -\frac{T}{12\pi}  \sum_{b=h,G^0,G^+,h^+,k^{++},S} n_b\bigg[M^3_b(\phi,T) - M^3_b(\phi)\bigg]. \label{daisy} 
\eea
In the above, $M^2_b(\phi,T) = M^2_b(\phi) + \Pi_b(T)$, with $\Pi_b(T)$ being the thermal correction to the mass of the $b$-th boson. One finds
\besub
\bea
\Pi_{h_0}(T) = \Pi_{G_0}(T) = \Pi_{G^+}(T)  &=& \frac{1}{12}\Big(\l_1 + \l_2 + \l_3 + 6\l_H + \frac{3}{4}(g^\prime)^2 + \frac{9}{4}g^2 + 3 y^2_t \Big)T^2, \\
\Pi_{h^+}(T) &=& \frac{1}{12}\Big(2\l_1 + \l_4 + \l_6 + 4\l_h + 3(g^\prime)^2 \nonumber \\
&&
 + 12 g^2_{BL} + \text{Tr}(Y_1^\dagger Y_1) \Big)T^2, \\
\Pi_{k^{++}}(T) &=& \frac{1}{12}\Big(2\l_2 + \l_5 + \l_6 + 4\l_k + 12(g^\prime)^2 + 12 g^2_{BL} \Big)T^2, \\
\Pi_{s_0}(T) = \Pi_{a_0}(T) &=& \frac{1}{12}\Big(2\l_3 + \l_4 + \l_5 + 3\l_S \nonumber \\
&&
 + 12 g^2_{BL} + \frac{1}{2}\text{Tr}(Y_2^\dagger Y_2) \Big)T^2.
\eea
\eesub
In addition, all couplings featuring in $V_T(\phi,T)$ and $V_{\text{Daisy}}(\phi,T)$ are determined at the energy scale $\phi$ through RG evolution. The total potential is given by $V_{\text{tot}}(\phi,T)=V_{\text{RG}}(\phi)+V_T(\phi,T) + V_{\text{Daisy}}(\phi,T)$. The thermal potential behaves as $\sim \phi^2 T^2$ for small field values implying that $\phi = 0$ is a minima at a nonzero temperature. In fact, $V_{\text{tot}}(\phi,T)$ is symmetric at high temperatures with a single minima at $\phi=0$. While the zero temperature RG-improved potential tends to dynamically break the $U(1)_{B-L}$ at $\phi = v_{BL}$, a symmetry-restoring effect comes from the $T\neq 0$ corrections leading to an interesting interplay. Therefore, $V_{\text{tot}}(\phi,T)$ develops an inflection point at a given temperature.
As further temperature lowers, a new minima ($\phi_{\text{min}}(T)$) appears thereby opening up the possibility of tunnelling. The tunnelling probability per unit 4-volume is given by~\cite{Linde1983},
\bea
\Gamma(T) \sim T^4 \bigg(\frac{S_E}{2\pi T}\bigg)^{3/2}e^{-\frac{S_E}{T}}.
\eea
Here, $S_E$ is the classical euclidean ``bounce" action in 3-dimensions calculated as
\bea
S_E = 4\pi \int_0^\infty dr~r^2 \bigg[\frac{1}{2}\Big( \frac{d \phi}{d r} \Big)^2 + V_{\text{total}}(\phi,T) \bigg].
\eea
The scalar field $\phi$ is derived by solving the classical field equation,
\bea
\frac{d^2 \phi}{d r^2} + \frac{2}{r}\frac{d\phi}{d r} - \frac{\partial V_{\text{total}}}{\partial \phi} = 0.
\eea
A critical temperature $T_c$ is identified through
\bea
V_{\text{tot}}(0,T_c) = V_{\text{tot}}(\phi_{\text{min}}(T_c),T_c),
\eea
An FOPT is deemed \emph{strong} if $\phi_{\text{min}}(T_c)/T_c > 1$~\cite{Bochkarev:1990gb,Quiros:1999jp}.
The bubble picture can be invoked to understand FOPT in analogy with the liquid-gas phase transition. In a sea of the false vacuum, it can be thought to be populated by bubbles containing the true vacuum. The bubbles grow in size as tunnelling progresses. At the nucleation temperature $T_n$, there is $\sim$ 1 bubble per unit Hubble volume~\cite{Moreno:1998bq}, i.e., $\Gamma(T_n) = H^4(T_n)$. The nucleation temperature can be computed using~\cite{McLerran:1990zh, Dine:1991ck, Moreno:1998bq},
\bea
\frac{S_E(T_n)}{T_n} = 140.
\eea
\begin{table}[t]
\centering
\small
\begin{tabular}{|c| c| c| c| c|}
\hline
Benchmark & $v_{BL}$(TeV) & $g_{BL}(0)$  & $\{y_{S1}(0),y_{S2}(0),y_{S3}(0)\}$ & $Y_1(0)$ \\
\hline \hline
BP1 &
16.2 &
0.49 &
$\{0.62,0.70,0.70\}$ &
$\begin{pmatrix}
0 & 0 & 0 \\
0.224 & 0.249 - 0.007 i & -0.239 - 0.006 i \\
-0.008 - 0.002 i & 0.043 & 0.038
\end{pmatrix}$ \\[6pt] 
\hline
BP2 &
22.0 & 0.54 & $\{0.70,0.76,0.76\}$ &
$\begin{pmatrix}
0 & 0 & 0 \\
0.261 & 0.291 - 0.008 i & -0.278 - 0.007 i \\
-0.009 - 0.0029i & 0.051 & 0.044
\end{pmatrix}$ \\[6pt]
\hline
BP3 &
40.0 & 0.52 & $\{0.72,0.77,0.77\}$ &
$\begin{pmatrix}
0 & 0 & 0 \\
0.352 & 0.392-0.011 i & -0.375 - 0.010 i \\
-0.013-0.004 i & 0.069 & 0.059
\end{pmatrix}$\\[6pt]
\hline
BP4 &
50.0 & 0.51 & $\{0.72,0.80,0.80\}$ &
$\begin{pmatrix}
0 & 0 & 0 \\
0.394 & 0.438-0.01 i & -0.420-0.01 i \\
-0.014 - 0.004 i & 0.077 & 0.066
\end{pmatrix}$\\
\hline
\end{tabular}
\caption{Benchmark points used in the numerical analysis.}
\label{tab:bp}
\end{table}
In order to test the strength of FOPTs in the present framework, we choose the following benchmark points shown in Table \ref{tab:bp}.
The values of the other parameters are fixed as,
\besub
\begin{gather}
\l_4(0)=0.1,~\l_5(0)=0.3,~\l_7(0)=0.5,~~
Y_2(0) = \begin{pmatrix}
1 & 1 & 1 \\
1 & 1 & 1 \\
1 & 1 & 1
\end{pmatrix} \times 10^{-2}, \\
\l_1(0) = \l_2(0) = \l_6(0) = 0.01.
\end{gather}
\eesub
The benchmarks are charaterised by $\mathcal{O}(0.1)$ values for the couplings that involve the scalar $S$.
The values taken for the various parameters are consistent with neutrino data as well as bounds from LFV and colliders. They are also compatible with the observed DM relic density and the LZ bound on direct detection cross sections as can be read from Table ~\ref{tab:bp_cosmo}. We add that we have handled the $J_{B/F}(x)$ functions numerically and refrained from taking simplifying high-$T$ expansions of the same. In addition, $S_E(T),~T_n$ are computed by fitting $V_{\text{tot}}(\phi,T)$ to an analytical form following Ref.~\cite{Adams:1993zs}.

\begin{figure}[H]
\centering
\includegraphics[scale=0.35]{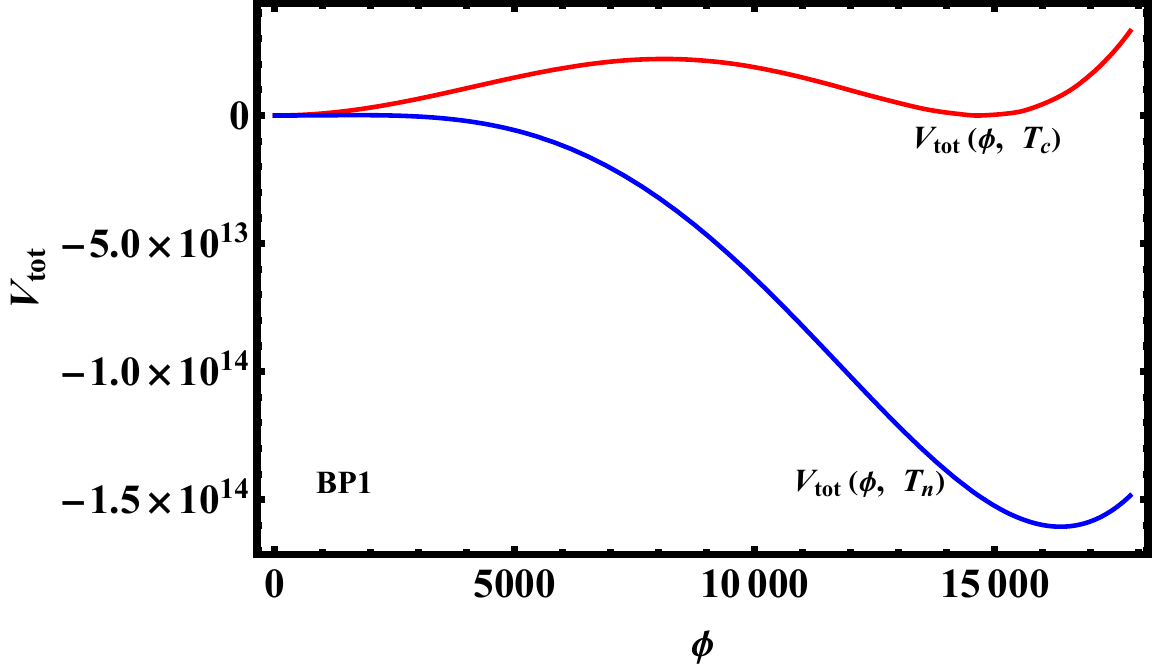}~~~
\includegraphics[scale=0.35]{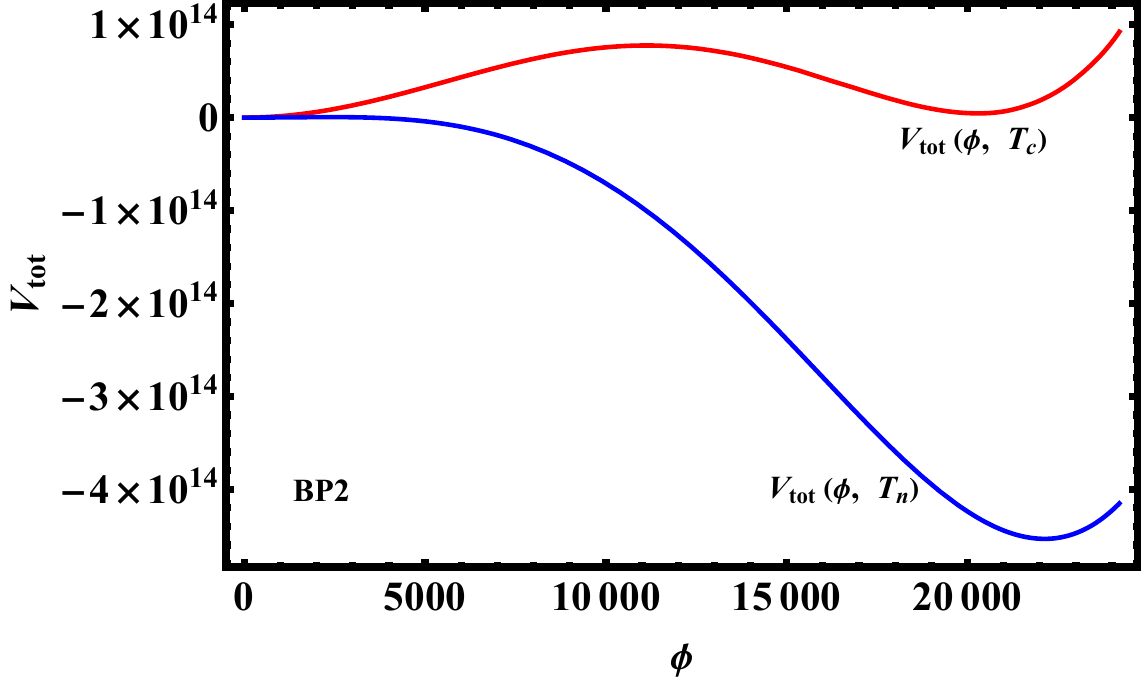}\\
\includegraphics[scale=0.35]{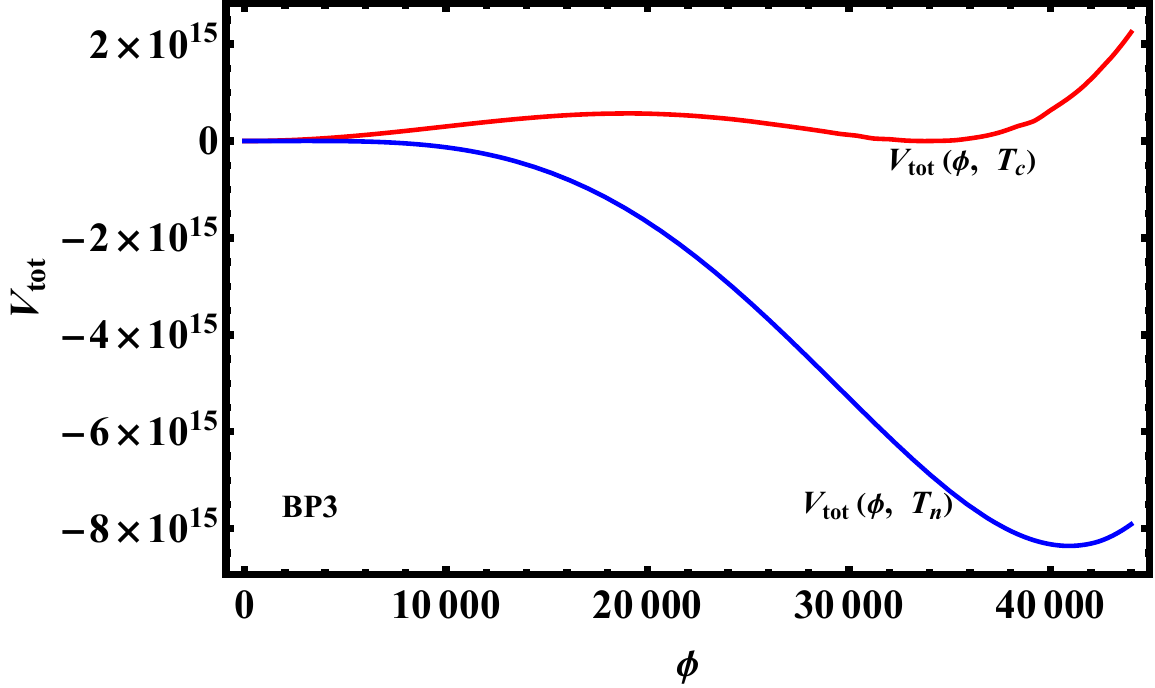}~~~
\includegraphics[scale=0.35]{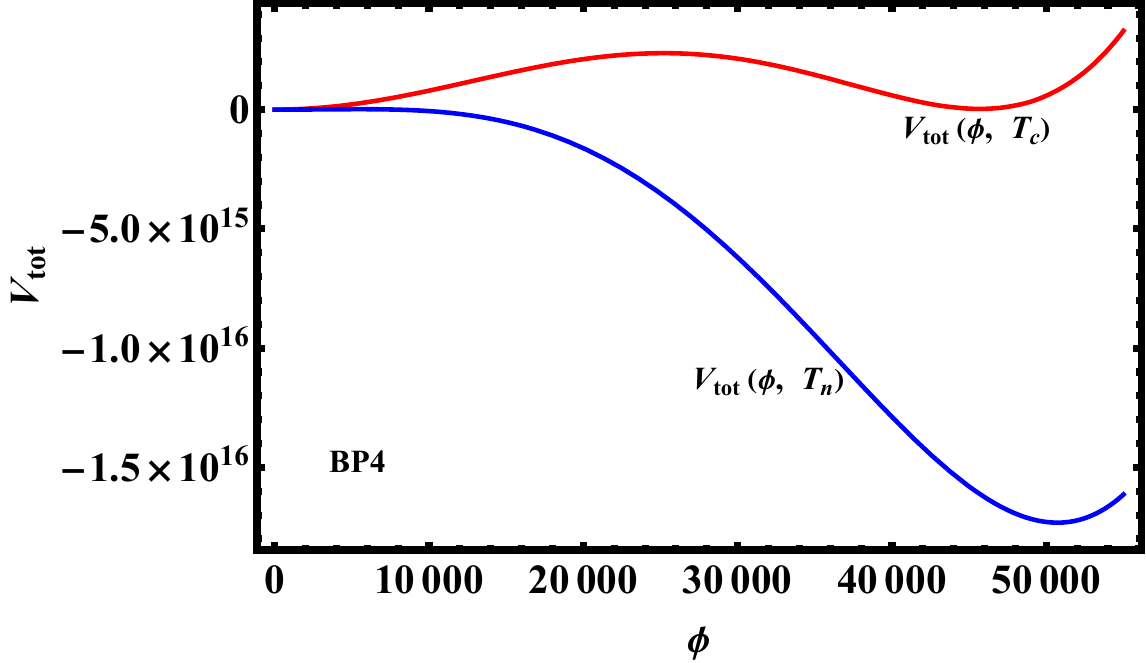}
\caption{The shape of $V_{\text{tot}}(\phi,T)$ at $T=T_c,T_n$ for the four chosen benchamrks.}
\label{fig:VT}
\end{figure}

The thermal potential profiles at $T=T_c,T_n$ for the chosen benchmarks is shown in Fig.~\ref{fig:VT}.
One inspects $\phi_{\text{min}}(T_c)/T_c > 1$ for all BP1-4 implying that the FOPT is \emph{strong} and this can be attributed to classical scale invariance. 
More importantly, the strength of the FOPT is mainly dictated by primarily the $g_{BL}(0)$ and the Yukawa couplings, i.e., as the $g_{BL}(0)$ increases the FOPT becomes weaker (see BP1 and BP4); along with that, for sufficiently large values of $g_{BL}(0)$, as the ratio $g_{BL}(0)/\mathrm{max}\{y_{S1}(0), y_{S2}(0), y_{S3}(0)\}$ decreases, the strength decreases (see BP2 and BP4). One also inspects that the ratio $T_c/T_n \in (2.75, 3.3)$ for all the BPs. A  $T_n$ significantly lower than $T_c$ implies that the phase transition is appropriately delayed.
Before going into the discussion of the gravitational waves due to the FOPT, we define two very important quantities that not only provide us a quantification of the FOPT property, but are linked directly to the estimation of the gravitational wave spectra. The first one signifies the apparent strength of the FOPT and can be expressed as,
\begin{align}
\alpha=\dfrac{1}{\rho_{\mathrm{rad}}(T_n)}\left[\Delta V_{\mathrm{tot}}-\dfrac{T}{4}\dfrac{d}{dT}(\Delta V_{\mathrm{tot}})\right]_{T=T_n},
\end{align}
where $\Delta V_{\mathrm{tot}}=V_{\mathrm{tot}}\vert_{\phi=\phi_f}-V_{\mathrm{tot}}\vert_{\phi=\phi_t}$, $\rho_{\mathrm{rad}}(T)=\pi^2 g_* T^4/30$ and $\phi_{f(t)}$ is the value of the field at the false (true) vacuum. The other important quantity gives us an estimate of the inverse duration of the FOPT and can be expressed as,
\begin{align}
\beta/H\approx T_n\left[\dfrac{d}{dT}\left(\dfrac{S_3}{T}\right)\right]_{T=T_n}.
\end{align}
 There are a few mechanisms through which gravitational waves can originate from FOPT, such as collision of bubble walls, the sound wave of the plasma, magnetohydrodynamic turbulence of the plasma, etc. The contribution of these sources depend on the distribution of the released latent heat, i.e., if more energy is released to the bubble walls then they tend to overcome the plasma friction and collide with each other making collision of bubble walls the dominant source whereas for weaker FOPTs more energy is released in the plasma and the friction inhibits the walls from colliding which makes the sound waves of the plasma the dominant source of gravitational waves.
On this note, the most important aspect to determine is the dominant source of gravitational wave background for this case. In order to achieve that, we calculate the leading order plasma friction $\Delta P_{\mathrm{LO}}$, the bubble wall velocity $v_w$ and the efficiency factor $\kappa$ which signifies the energy distribution.
The leading order friction term can be expressed as~\cite{Bodeker:2009qy,Ellis:2019oqb},
\begin{align}
    \Delta P_{\mathrm{LO}} = \frac{\Delta m^2 T^2}{24},
\end{align}
where 
\begin{align}
    \Delta m^2 = \sum_i c_i N_i \Delta m_i^2,
\end{align}
with $\Delta m_i^2$ is the difference of the squared mass of the $i$-th particle species in true and false vacuum, $N_i$ is the degree of freedom of the $i$-th particle species and $c_i = 1 (1/2)$ for bosons (fermions). Now the plasma friction strength can be calculated as.
\begin{align}
    \alpha_{\infty} = \frac{\Delta P_{\mathrm{LO}}}{\rho_R},
\end{align}
where $\alpha = \alpha_{\infty}$ denotes the condition where the driving force on the vacuum wall is larger than the leading order friction. For $\alpha \gg \alpha_{\infty}$ leads to the creation of runaway bubbles and bubble wall collision dominates, whereas for $\alpha \ll \alpha_{\infty}$ the bubble wall velocity is much less than the speed of light in most cases, and sound waves are the dominant source. However, we found that for the benchmark cases that we consider, $\alpha_{\infty}$ is marginally less than $\alpha$, which leads to the scenario where in all cases, the gravitational waves due to both the bubble wall collision and the sound wave have to be considered.\footnote{It is to be noted that although several studies have investigated the GW spectra from the magneto-hydrodynamic turbulence of plasma surrounding the bubble wall~\cite{RoperPol:2019wvy, Kahniashvili:2020jgm, RoperPol:2021xnd, Auclair:2022jod}, due to the existence of uncertainties in the spectral shape, we do not consider this source~\cite{Caprini:2015zlo}.}

In order to calculate the contribution of the sound wave to the gravitational waves, we first lay the foundation with the calculation of the bubble wall velocity and the efficiency factor for the sound waves. To calculate the bubble wall velocity, we follow the prescription of Ref.~\cite{Ellis:2022lft}, and use the approximate expression~\cite{Lewicki:2021pgr},
\begin{align}
  v_w =  \begin{cases}
       \sqrt{\frac{\Delta V}{\alpha \rho_R}}, \text{~for~} \sqrt{\frac{\Delta V}{\alpha \rho_R}}<v_J,\\
       1, \text{~for~} \sqrt{\frac{\Delta V}{\alpha \rho_R}}>v_J,
    \end{cases}
\end{align}
where $v_J$ is the Chapman-Jouguet velocity and can be expressed as~\cite{Steinhardt:1981ct,Kamionkowski:1993fg,Espinosa:2010hh},
\begin{align}
    v_J = \frac{1}{\sqrt{3}}\frac{1+\sqrt{3\alpha^2 + 2\alpha}}{1+\alpha}.
\end{align}
Using the above expression once the bubble wall velocity is obtained, one can then obtain the efficiency factor of the sound waves using the numerical fitting obtained in Ref.~\cite{Espinosa:2010hh}. The efficiency factor depends on the wall velocity and can be expressed as,
\begin{align}
    \kappa_{\mathrm{sw}} \approx 
    \begin{cases}
        \dfrac{c_s^{11/5}\kappa_A\kappa_B}{(c_s^{11/5}-v_w^{11/5})\kappa_B+v_w c_s^{6/5}\kappa_A}\text{~for~}v_w\lesssim c_s,\\
        \kappa_B + (v_w-c_s)\delta_{\kappa}+\dfrac{(v_w-c_s)^3}{(v_J-c_s)^3}[\kappa_C-\kappa_B-(v_J-c_s)\delta_{\kappa}]\text{~for~}c_s<v_w<v_J,\\
        \dfrac{(v_J-1)^3v_J^{5/2}v_w^{-5/2}\kappa_C\kappa_D}{[(v_J-1)^3-(v_w-1)^3]v_J^{5/2}\kappa_C+(v_w-1)^3\kappa_D}\text{~for~}v_w\gtrsim v_J,
    \end{cases}
    \label{eq:kappa1}
\end{align}
where,
\begin{align}
    \kappa_A &\approx v_w^{6/5}\dfrac{6.9\alpha}{1.36-0.037\sqrt{\alpha}+\alpha},\\
    \kappa_B &\approx \dfrac{\alpha^{2/5}}{0.017+(0.997+\alpha)^{2/5}},\\
    \kappa_C &\approx \dfrac{\sqrt{\alpha}}{0.135+\sqrt{0.98+\alpha}},\\
    \kappa_D &\approx \dfrac{\alpha}{0.73+0.083\sqrt{\alpha}+\alpha},\\
    \delta_{\kappa} &\approx -0.9\log\dfrac{\sqrt{\alpha}}{1+\sqrt{\alpha}},
\end{align}
and $c_s = 1/\sqrt{3}$ is the speed of sound wave in plasma. It is to be noted that the first, second, and the third cases in Eq.~\eqref{eq:kappa1} are sub-sonic deflagration, supersonic deflagration, and detonation respectively.

\begin{table}[t]
\centering
\label{tab:plain6}
\rowcolors{2}{gray!20}{white}
\begin{tabular}{|c| c| c| c| c| c| c| c| c| c| c|}
\hline
 \rowcolor{gray!50}
Benchmark & $\Omega_{N_1}h^2$ & $\sigma_{SI}(\text{cm}^2)$ & $T_c$(TeV)
 & $\phi_c/T_c$ & $T_n$(TeV) & $v_w$ & $\kappa_{sw}$ & $\alpha$ & $\beta/H$ & $\alpha_{\infty}$  \\
\hline \hline
BP1  & 0.117 & $9.91 \times 10^{-47}$  & 7.8 & 1.91 & 2.73 & 0.752 & 0.455 & 0.141 & 198 & 0.082 \\ \hline
BP2 & 0.111 & $2.91 \times 10^{-47}$ & 17.65 & 1.15 & 6.25 & 0.449 & 0.1938 & 0.079 & 189 & 0.034 \\ \hline
BP3 & 0.113 & $2.66 \times 10^{-48}$ & 29.5 & 1.20 & 9.02 & 0.592 & 0.392 & 0.096 & 194 & 0.053 \\ \hline
BP4 & 0.103 & $1.09 \times 10^{-48}$ & 25.6 & 1.79 & 9.27 & 0.713 & 0.482 & 0.126 & 208 & 0.075 \\
\hline
\end{tabular}
\caption{Parameters pertaining to DM, FOPT and GWs corresponding to the chosen benchmarks}
\label{tab:bp_cosmo}
\end{table}

Finally, the SGWB spectra due to sound waves can be expressed as~\cite{Hindmarsh:2013xza, Hindmarsh:2015qta, Hindmarsh:2017gnf,Hindmarsh:2020hop},
\begin{align}
\Omega_{\mathrm{sw}}h^2 = 4.13\times 10^{-7}(R_*H_*)\left(1-\dfrac{1}{\sqrt{1+2\tau_{\mathrm{sw}}H_*}}\right)\left(\dfrac{\kappa_{\mathrm{sw}}\alpha}{1+\alpha}\right)^2\left(\dfrac{100}{g_*}\right)^{1/3}S_{\mathrm{sw}}(f),
\end{align}
where,
$S_{\mathrm{sw}}$ contains the information regarding the frequency dependence of the SGWB spectrum and can be expressed as,
\begin{align}
S_{\mathrm{sw}}(f) = \left(\dfrac{f}{f_{\mathrm{sw}}}\right)^3\left(\dfrac{4}{7}+\dfrac{3}{7}\left(\dfrac{f}{f_{\mathrm{sw}}}\right)^2\right)^{-7/2},
\end{align}
the peak frequency of the spectrum is given by,
\begin{align}
f_{\mathrm{sw}} = 2.6\times 10^{-5}(R_*H_*)^{-1}\left(\dfrac{T_{n}}{100~\mathrm{GeV}}\right)\left(\dfrac{g_*}{100}\right)^{1/6}\mathrm{~Hz},
\end{align}
and the other relevant quantities are defined as~\cite{Ellis:2018mja,Ellis:2019oqb,Hindmarsh:2017gnf,Ellis:2020awk,Guo:2020grp},
\begin{align}
\tau_{\mathrm{sw}}H_* &= \dfrac{R_*H_*}{U_f},\\
U_f &= \sqrt{\dfrac{3}{4}\dfrac{\alpha}{1+\alpha}\kappa_{\mathrm{sw}}},\\
R_*H_* &= (8\pi)^{1/3}\mathrm{max}(v_w,~c_s)\left(\dfrac{\beta}{H}\right)^{-1}.
\end{align}

On the other hand, to estimate the contribution of the bubble wall collision to the gravitational waves, one must first consider the efficiency factor of the energy released into the bubble walls. It has been shown that the efficiency factor can be expressed as~\cite{Caprini:2015zlo},
\begin{align}
    \kappa_{\mathrm{coll}} \approx \dfrac{\alpha - \alpha_{\infty}}{\alpha}.
\end{align}
Hence the SGWB spectra due to the bubble wall collision today can be expressed as~\cite{Cutting:2018tjt,Ellis:2019oqb,Lewicki:2022pdb},
\begin{align}
    \Omega_{\mathrm{coll}}h^2 = 4\times 10^{-7}(R_*H_*)^2\left(\dfrac{100}{g_*}\right)^{1/3}\left(\dfrac{\kappa_{\mathrm{coll}}\alpha}{1+\alpha}\right)^2S_{\mathrm{coll}}(f),
\end{align}
where $S_{\mathrm{coll}}$ contains the information regarding the frequency dependence of the SGWB spectra and can be expressed as,
\begin{align}
    S_{\mathrm{coll}}(f) = \left(\dfrac{f}{f_{\mathrm{coll}}}\right)^3\left(1+2\left(\dfrac{f}{f_{\mathrm{coll}}}\right)^{2.07}\right)^{-2.18},
\end{align}
and the peak frequency of the spectrum can be expressed as
\begin{align}
    f_{\mathrm{coll}} = 2.88\times 10^{-6}\left(\dfrac{1}{v_w}\right)\left(\dfrac{\beta}{H}\right)\left(\dfrac{T_n}{100\mathrm{~GeV}}\right)\left(\dfrac{g_*}{100}\right)^{1/6}\mathrm{~Hz}.
\end{align}
It is worth mentioning here that although the SGWB is calculated at the reheating temperature, in this case since $\alpha\ll 1$ for all the cases that we consider, the reheating temperature $T_{\mathrm{reh}}\approx T_{\mathrm{n}}$.
Finally, the combined SGWB can be expressed as,
\begin{align}
    \Omega_{\mathrm{GW}}h^2 = \Omega_{\mathrm{sw}}h^2 + \Omega_{\mathrm{coll}}h^2.
\end{align}

\begin{figure}[t]
\centering
\includegraphics[scale=0.8]{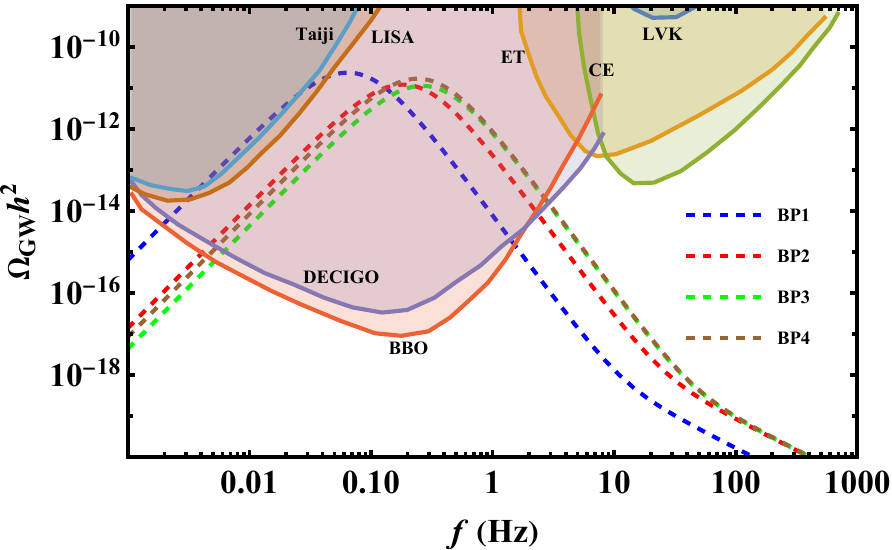}
\caption{The gravitational wave signatures arising from the FOPT for the benchmark parameters considered. The projected sensitivity of current and upcoming GW experiments has been shown.}
\label{gw}
\end{figure}
We show the variation of $\Omega_{\text{GW}}h^2$ versus the frequency $f$ in Fig.~\ref{gw}. The sensitivities of different future space-based GW interferometers such as LISA~\cite{LISA:2017pwj}, BBO~\cite{Harry:2006fi}, DECIGO~\cite{Kawamura:2011zz}, Taiji~\cite{Ruan:2018tsw}, ET~\cite{Punturo:2010zz}, CE~\cite{Reitze:2019iox} and LVK (LIGO-Virgo-KAGRA)~\cite{LIGOScientific:2014pky, VIRGO:2012dcp, Somiya:2011np} are also
shown in Fig.~\ref{gw}. The GW spectra for the chosen BPs peak in the frequency range $\mathcal{O}(0.01)-\mathcal{O}(0.1)$ Hz with the peak mildly shifting towards the right as $v_{BL}$ increases. This implies that while BP2-4 are observable at BBO and DECIGO, BP1 is within the purview of LISA.

\section{Conclusions}
\label{conclu}
In this work, we have proposed a scale invariant version of the Zee-Babu model governed by an $U(1)_{B-L}$ gauge symmetry. The Zee-Babu framework also demands lepton number violation in the Lagrangian in order to address Majorana masses for the SM neutrinos, much like other seesaw scenarios. We argue that such a violation of a global $U(1)_L$ can be a consequence of a broken local $U(1)_{B-L}$. Anomaly cancellation necessitates including three RHNs which are chosen here to carry an $\mathbb{Z}_2$ charge so that the lightest among them can be a potential DM candidate. The breaking of the $U(1)_{B-L}$ symmetry is catalysed by RG effects leading to a dynamically generated scale $v_{BL}$, which in turn triggers EWSB. 
The crucial scalar trilinear coupling featuring in the minimal Zee-Babu set-up is now seen to get generated from a scalar quartic interaction via symmetry breaking. A key finding of this study is that the light neutrino mass scale is dictated by the symmetry breaking scale $v_{BL}$, given the classically scale invariant nature of the framework. In addition, LFV processes and the DM observables of relic density and direct detection cross section are also controlled by the same $v_{BL}$. We have adopted the Casas-Ibarra parameterisation to fit the neutrino oscillation data. Demanding that the lightest RHN alone accounts for the observed relic density and predicts an appropriately low DM-nucleon scattering rate constrains the symmetry breaking scale in the 10 TeV $\leq v_{BL} \leq$ 55 TeV range. 
We have also studied FOPTs in the aforementioned range by incorporating $T\neq 0$ corrections into the scalar potential. It is also demonstrated through representative parameter points that a \emph{strong} FOPT can occur concomitantly with explaining the observed neutrino data and DM relic abundance while satisfying the bounds from LFV and DM direct detection. The corresponding stochastic GW spectra are mainly shaped by contributions from sound waves and bubble collisions, with the sound wave component generally dominating due to efficient energy transfer in relativistic fluids. We further report that GW spectra for the chosen parameter points can be probed using the future space-based detectors LISA and BBO. 
Overall, an $U(1)_{B-L}$-extended Zee-Babu model emerges as an attractive framework to test classical scale invariance through the window of neutrino mass and WIMP DM when the gauge symmetry is radiatively broken at an $\mathcal{O}$(10 TeV) scale. Moreover, the model also leads to potentially detectable stochastic GW signatures.

\paragraph*{Acknowledgements\,:} 
UKD acknowledges support from the Anusandhan National Research Foundation (ANRF), Government of India under Grant Reference No.~CRG/2023/003769.

\appendix
\section*{Appendix}
\label{s:appen}
The one-loop $\beta$-functions for this model are given below. 
\besub
\bea
16 \pi^2\beta_{\l_H} &=& 24 \l^2_H + \l^2_1 + \l^2_2 + \l_3^2 - \l_H\big(3 (g^\prime)^2 + 9 g^2 - 12 y^2_t \big) \nonumber \\
&&
+ \frac{3}{8}(g^\prime)^4 + \frac{9}{8} g^4 + \frac{3}{4}(g^\prime)^2 g^2
- 6 y^4_t, \\
16 \pi^2\beta_{\l_h} &=& 20 \l^2_h + 2 \l^2_1 + \l_4^2 + 
\l_6^2 + 2 \l_7^2 - \l_h \big( 12 (g^\prime)^2 + 48 g^2_{BL} \big) \nonumber \\
&&
 + 6 (g^\prime)^4 + 48 (g^\prime)^2 g^2_{BL} + 96 g^4_{BL} , \\
16 \pi^2\beta_{\l_k} &=& 20 \l^2_k + 2 \l^2_2 + \l_5^2 + \l_6^2
  - \l_k \big( 48(g^\prime)^2 + 48 g^2_{BL} \big) \nonumber \\
&&  
   + 96 (g^\prime)^4 + 192 (g^\prime)^2 g^2_{BL} + 96 g^4_{BL},\\
16 \pi^2\beta_{\l_S} &=& 20 \l_S^2 + 2 \l_3^2 + \l_4^2 + \l_5^2 
-\l_S \big( 48 g^2_{BL} - 8 \text{Tr}[Y_S^\dagger Y_S] \big) \nonumber \\
&&
 + 96 g^4_{BL} - \text{Tr}[Y_S^\dagger Y_S Y_S^\dagger Y_S], \\
16 \pi^2\beta_{\l_1} &=& 4 \l_1^2 + 2\l_3 \l_4 + 2 \l_2 \l_6+ 12 \l_1 \l_H + 8 \l_1 \l_h  \nonumber \\
&&
 - \l_1 \Big( \frac{15}{2}(g^\prime)^2 + \frac{9}{2}g^2 + 24 g^2_{BL} - 6 y^2_t \Big) + 3 (g^\prime)^4, \\
16 \pi^2\beta_{\l_2} &=& 4 \l_2^2 + 2\l_3 \l_5 + 2 \l_1 \l_6 + 12 \l_2 \l_H + 8 \l_2 \l_k \nonumber \\
&&
 - \l_2 \Big( \frac{51}{2}(g^\prime)^2 + \frac{9}{2}g^2 + 24 g^2_{BL} - 6 y^2_t \Big) + 12 (g^\prime)^4, \\
16 \pi^2\beta_{\l_3} &=& 4 \l_3^2 + 2 \l_1 \l_4 + 2 \l_2 \l_5 + 12 \l_3 \l_H + 8 \l_3 \l_S \nonumber \\
&&
 - \l_3\Big(\frac{3}{2}(g^\prime)^2 + \frac{9}{2}g^2 + 24 g^2_{BL} - 6 y^2_t - \text{Tr}[Y^\dagger_S Y_S] \Big), \\
16 \pi^2\beta_{\l_4} &=& 4 \l_4^2 + 4 \l_1 \l_3 + 2 \l_5 \l_6 + 8 \l_7^2 + 8 \l_4 \l_h + 8 \l_4 \l_S \nonumber \\
&&
 - \l_4 \big(6(g^\prime)^2 + 48 g^2_{BL} - \text{Tr}[Y^\dagger_S Y_S] \big) + 192 g^4_{BL}, \\
16 \pi^2\beta_{\l_5} &=& 4 \l_5^2 + 4 \l_2 \l_3 + 2 \l_4 \l_6 + 4 \l_7^2 + 8 \l_5 \l_k + 8 \l_5 \l_S  \nonumber \\
&&
 - \l_5\big(24(g^\prime)^2 + 48 g^2_{BL} - \text{Tr}[Y^\dagger_S Y_S] \big) + 192 g^4_{BL}, \\
16 \pi^2\beta_{\l_6} &=& 4 \l_6^2 + 4 \l_1 \l_2 + 2\l_4 \l_5 + 8 \l_7^2 + 8 \l_6 \l_h + 8 \l_6 \l_k \nonumber \\
&&
- \l_6 \big(30 (g^\prime)^2 + 48 g^2_{BL} \big)
 + 48 (g^\prime)^4 + 192 (g^\prime)^2 g^2_{BL} + 192 (g^\prime)^4, \\
16 \pi^2\beta_{\l_7} &=& \l_7\big(4 \l_4 + 2 \l_5 + 4 \l_6 + 4 \l_h - 12 (g^\prime)^2 - 32 g^2_{BL} + \frac{1}{2} \text{Tr}[Y^\dagger_S Y_S] \big).
\eea
\eesub

\besub
\bea
16 \pi^2 \beta_{y_t} &=& y_t \Big( \frac{9}{2}y^2_t - \frac{17}{12}(g^\prime)^2 - \frac{9}{4}g^2 - 8 g^2_s \Big), \\
16 \pi^2 \beta_{Y_S} &=&  Y_S Y_S^\dagger Y_S + \frac{1}{2} Y_S\text{Tr}[Y_S^\dagger Y_S] - 6 g^2_{BL}Y_S. 
\eea
\eesub

\besub
\bea
16 \pi^2\beta_{g^\prime} &=& \frac{17}{2} (g^\prime)^3,\\
16 \pi^2\beta_{g_{BL}} &=& \frac{44}{3} (g_{BL})^3,\\
16 \pi^2\beta_{g} &=& -\frac{19}{6}g^3,\\
16 \pi^2\beta_{g_s} &=& - 7 g^3_s.
\eea
\eesub


\bibliographystyle{JHEP}
\bibliography{refFB} 
\end{document}